\input phyzzx
\overfullrule=0pt
\tolerance=5000
\twelvepoint
\def\tilde{\widetilde}

\pubnum{IASSNS-HEP-98/43}
\date{hep-th/9806016}
\titlepage
\title{A Modular Invariant Partition Function for the Fivebrane}
\vglue-.25in
\author{ Louise Dolan \foot{Research supported in part by the U.S. Department
of Energy under Grant No. DE-FG 05-85ER40219/Task A. 
Email: dolan@physics.unc.edu}} 
\address{Department of Physics and Astronomy
\break University of North Carolina
\break Chapel Hill
\break North Carolina 27599-3255}
\author{Chiara R. Nappi\foot{Research supported in part by
the Ambrose Monell Foundation. Email: nappi@ias.edu}}
\medskip
\address{School of Natural Sciences
\break Institute for Advanced Study
\break Olden Lane
\break Princeton, NJ 08540}
\bigskip

\parskip=6pt

\abstract{We compute an $SL(6,{\cal Z})$ invariant partition function
for the chiral two-form of the M theory fivebrane
compactified on the six-torus $T^6$. From a manifestly
$SL(5,{\cal Z})$ invariant formalism, we prove that the partition
function has  an additional
$SL(2,{\cal Z})$ symmetry. The combination of these two symmetries
ensures $SL(6,{\cal Z})$ invariance. Thus, whether or not a fully
covariant Lagrangian is  available, the fivebrane on the six-torus has a
consistent quantum theory.}

\REF\Witten{E. Witten, ``Five-brane Effective Action in M-theory,''
J. Geom. Phys. {\bf 22} (1997) 103; hep-th 9610234.}

\REF\Perry{M. Perry and J. H. Schwarz, ``Interacting Chiral Gauge
Fields in Six Dimensions and Born-Infeld Theory','' 
Nucl. Phys. {\bf B489} (1997) 47; hep-th/9611065.}

\REF\Schwarz{ John H. Schwarz, ``Coupling a Self-dual Tensor to
Gravity in Six Dimensions,''Phys. Lett. {\bf B395} (1997) 191; 
hep-th/ 9701008.}

\REF\PST{P. Pasti, D. Sorokin and M. Tonin, ``Covariant Action for a
D=11 Five-Brane with Chiral Field,'' Phys. Lett. {\bf B398} (1997) 41;
hep-th 9701037; ``On Lorentz
Invariant Actions for Chiral P-Forms,'' Phys.
Rev. {\bf D52} (1995) 4277; hep-th/9711100.}
 
\REF\Aganagic{ M. Aganagic, J. Park, C. Popescu and J. H. Schwarz,
``Worldvolume Action of the M-theory Fivebrane,'' Nucl. Phys. {\bf
B496} (1997) 191; hep-th/9701166.} 

\REF\Howe{ P.S. Howe, E. Sezgin and P.C. West, ``Covariant Field
Equations of the M-theory Five-brane, Nucl.Phys. {\bf B496} (1997)
191.}

\REF\Bandos{ I. Bandos, K. Lechner, A. Nurmagambetov, P. Pasti,
D. Sorokin and M. Tonin, ``On the Equivalence of Different
Formulations of M-theory Fivebrane,'' Phys. Lett. {\bf 408B} (1997)
135.}

\REF\Nilsson{M. Cederwall, B. Nilsson, P. Sundell, ``An Action for
the super-5-brane in $D=11$ Supergravity'', J. High Energy Physics
04 (1998) 7; hep-th/9712059.} 

\REF\Berg{E. Bergshoeff, D. Sorokin and P.K. Townsend, ``The M-5brane
Hamiltonian,'' hep-th/9805065.}   

\REF\Mumford {D. Mumford, {\it Tata Lectures on Theta} vols. I and II,
Boston: Birkhauser 1983; L. Alvarez-Gaume, G. Moore, C. Vafa,
``Theta Functions, Modular Invariance, and Strings'', Comm. Math. Phys.
{\bf 106} (1986) 1-40.}

\REF\hop {M. Hopkins and I. M. Singer, to appear}

\REF\Green{M.B. Green, J. H. Schwarz and E. Witten, 
$\underline{\rm Superstring\, Theory}$,
vol. I and II, Cambridge University Press: Cambridge, U.K. 1987.}

\REF\Coxeter{ H.S.M. Coxeter and W.O.J. Moser,
$\underline{\rm Generators\, and\, Relations\, for\, Discrete\, Groups}$,
Springer-Verlag: New York 1980.}

\chapter {Introduction}

The physical degrees of freedom of the 6d world volume theory of the M
theory fivebrane consist of an N=(2,0) tensor supermultiplet. This
multiplet contains a chiral two-form  
$B_{MN}$ with a self-dual three-form field strength 
$H_{LMN}= \partial_LB_{MN} + \partial_MB_{NL} + \partial_NB_{LM}$ 
where $1\le L,M,N \le 6$.
 
Many attempts
have been made recently to write a manifestly covariant
six-dimensional action for $H_{LMN}$ [\Perry 
--\Berg]. The difficulty is related to the self-duality
of the three-form field strength and is analogous to 
the  well-known problem of writing down covariant
actions for theories with chiral bosons. We adopt in this paper a
different approach. Rather than considering the Lagrangian, 
we ask if there is a modular invariant partition function. 

It is known there is no modular invariant partition function for a 
single chiral field in two dimensions. This is the reason, a posteriori,
that one cannot write a  covariant Lagrangian for such a field.
Such a Lagrangian, if it were to exist, could then be quantized
on a Riemann surface of genus $g$ and would yield results that depended only
on the metric of the Riemann surface in a modular invariant way.
Instead, it is known that a chiral scalar on a Riemann surface of
genus $g$ has $2^{2g}$ candidate partition functions.
The situation is expected to be similar for the partition function of
the M-theory fivebrane [\Witten].

Nonetheless in this paper  we compute 
the partition function $Z$ for a free self-dual $H_{LMN}$ field strength
on $T^6$ and show that  it is invariant 
under the $SL(6,{\cal Z})$ mapping class group of $T^6$. 
We circumvent the lack of a fully covariant Lagrangian by 
writing the self-dual three-form in six dimensions as  
an anti-symmetric three-form in five dimensions, which is no longer self-dual. 
In this formalism, the partition function is automatically $SL(5,{\cal Z})$
invariant. 
We show it has an additional
$SL(2,{\cal Z})$ invariance and from this prove that  
$Z$ is an example of an $SL(6,{\cal Z})$ automorphic form, 
even though only $SL(5,{\cal Z})$ symmetry is explicit.
It is given in (5.4) together with (3.11). 

The reason this partition function avoids the problems of the
chiral boson and manages to be modular invariant is that
we are compactifying on $T^2\times T^4$. From the point of view of $T^2$,
the three degrees of freedom of the two-form potential (which is the
(3,1) representation of the Spin(4)$\cong SU(2)\times SU(2)$ little
group for 6d, $N=(2,0)$ massless states)
behave like three 2d-massive scalars, therefore mimicking the situation  
of three non-chiral bosons.
This would not have happened had we compactified on $T^2\times {\bf CP2}$
for example. This is the way our  calculation and hence our result are
special to the compactification on $T^6$.

It was pointed out in the introduction to
[\Witten] that for a fivebrane on $\Sigma\times
{\bf CP}^2$ (with $\Sigma$ a Riemann surface) the chiral two-form
partition function depends on a spin structure on $\Sigma$. 
Our result shows
that in the case of $T^6$, there is no such spin structure dependence.
We understand
that, by relating the description of [\Witten] to the Kervaire
invariant,
M. Hopkins and I. M. Singer have determined when there is or is not a
dependence
of the chiral two-form partition function on the spin structure [\hop].
Notice  that 
for the partition function of a chiral two-form to depend on a spin
structure does not violate any symmetry of M-theory, since M-theory anyway
has fermions that require a spin structure. 
               
Although our result is restricted to the case of compactification on
the torus, we think it is interesting to exhibit an $SL(6, {\cal Z})$ 
modular invariant partition function for the M5-brane, {\it i.e.} a
quantum theory of the free 6d self-dual two-form with symmetry analogous to
the modular invariance of consistent interacting strings.  
In sect.3 we calculate the zero-mode contribution to $Z$, and in
sect.'s 4,5 we compute the oscillator trace and show how 
the $SL(2,{\cal Z})$  anomalies cancel. 

Appendix A describes an $SL(5, {\cal Z})$ invariant regularization
of  the vacuum energy in the trace over oscillators.
Appendix B reviews how
`2-d massive' part of our partition function is  $SL(2, {\cal Z})$ invariant,
an important ingredient of our proof.
Finally,  Appendix C deals with 
the generators of the mapping class groups
$SL(n,{\cal Z})$ for the n-torus and shows how $SL(2, {\cal Z})$ and 
$SL(5, {\cal Z})$ combine to give $SL(6, {\cal Z})$ invariance.

\chapter {Notation and Strategy}

The 6d chiral two-form $B_{MN}$ has a self-dual field-strength which satisfies
$$H_{LMN}(\vec\theta,\theta^6)= {1\over{6\sqrt{-G}}}G_{LL'}G_{MM'}G_{NN'}
\epsilon^{L'M'N'RST}H_{RST}(\vec\theta,\theta^6) \eqn\A $$
where $G_{MN}$ is the 6d metric.
With use of this equation of motion it is possible [\Schwarz] 
to eliminate the components $H_{6mn}$
in terms of the other components $H_{lmn}$ with $l,m,n
=1...5$ as follows
$$H_{6mn} = -{G^{6l}\over G^{66}} H_{lmn}
+ {1\over{6\sqrt{-G}G^{66}}}\epsilon_{mnlrs}H^{lrs}, \eqn\DEP $$
where now indices are raised with the 5d-metric $G_5^{mn}$ and 
$\epsilon_{12345} \equiv G_5\epsilon^{12345} = G_5$.
 
The Hamiltonian and momenta of the $H_{lmn}$ 
can be written in a fully 5d covariant way:
$${\cal H} = {1\over 12}{\int_0}^{2\pi}
d\theta^1...d\theta^5{\sqrt G_5}{G_5}^{ll'}
{G_5}^{mm'}{G_5}^{nn'} H_{lmn}(\vec\theta,\theta^6)\, 
H_{l'm'n'}(\vec\theta,\theta^6) \eqn\Ham$$
$$P_l = -{1\over 24}{\int_0}^{2\pi}
d\theta^1...d\theta^5 \epsilon^{rsumn} H_{umn}(\vec\theta,\theta^6)\,
H_{lrs}(\vec\theta,\theta^6)\eqn\Pel$$
with $1\le l,m,n,r,s,u\le 5$.

A general metric on $T^6$ is a function of 21 parameters and 
can be represented by the line element
$$\eqalign{ds^2 = & {R_1}^2(d\theta^1 -\alpha d\theta^6)^2 +
{R_6}^2(d\theta^6)^2 \cr
+&\sum_{i,j=2...5}g_{ij}(d\theta^i -
\beta^id\theta^1-\gamma^id\theta^6)
(d\theta^j - \beta^jd\theta^1 - \gamma^jd\theta^6) \cr}\eqn\B$$
where $0\le\theta^I\le 2\pi$, $1\le I\le 6$ and 
we single out  directions 1 and 6. The latter is our time direction.
The 21 parameters are as follows: $R_1$ and $R_6$ are the radii for 
directions 1 and 6,  
$g_{ij}$ is a 4d metric, $\beta^i,\gamma^j$ are the angles between
directions  1 and $i$, and between 6 and $j$, respectively; and  
$\alpha$ is related to the angle between 1 and 6, see (2.6).
The metric can be read off from the line element above to be  
$$\eqalign{ &{G}_{ij}= g_{ij}\,;
\qquad \hskip20pt G_{11}= {R_1}^2 + g_{ij}\beta^i\beta^j\cr 
&G_{i1}= -g_{ij}\beta^j\,;
\qquad G_{66}= {R_6}^2 + \alpha^2R_1^2 +g_{ij}\gamma^i\gamma^j\cr  
&G_{i6}= -g_{ij}\gamma^j\,;
\qquad G_{16}= -\alpha R_1^2 + g_{ij}\beta^i\gamma^j\,. \cr } \eqn\Gsix$$ 
Its inverse is 
$$\eqalign{G^{ij}=& {g}^{ij} +   {{\beta^i\beta^j}\over R_1^2}
+ {1  \over R_6^2}(\alpha \beta^i + \gamma^i)(\alpha \beta^j +
\gamma^j)\,;\hskip10pt
\qquad G^{11} = {1\over R_1^2}  + {\alpha^2 \over R_6^2}\cr
G^{1i}=& {\beta^i\over R_1^2} + {\alpha  \over R_6^2}(\alpha \beta^i +
\gamma^i)\,;
\qquad \hskip100pt G^{66} = {1\over R_6^2}\cr
G^{i6} =&{{(\alpha \beta^i + \gamma^i)} \over R_6^2}\,; 
\hskip150pt \qquad
G^{16} = {\alpha \over R_6^2}\,.\cr}\eqn\Gsixinv$$
The 5-dimensional inverse is  
$$\eqalign{ {G_5}^{ij}=& g^{ij} + {{\beta^i\beta^j}\over R_1^2}\,;\qquad
{G_5}^{i1}={\beta^i\over R_1^2}\,;\qquad
{G_5}^{11}= {1\over R_1^2}\cr} \eqn\Gfiveinv$$
where ${g}^{ij}$ is the 4d inverse of $g_{ij}$. Useful formulae are 
$G_5 = {R_1}^2 g, $ where $G_5 = {\rm det} G_{mn}$ 
and $g = {\rm det} g_{ij}$. Also $G=\det G_{IJ}$. 

We want to compute the  partition function on the twisted torus with metric
\Gsix.  As in string theory [\Green],  the partition function 
is given in terms of the Hamiltonian and momenta  by 
$$Z= tr(e^{-t{\cal H} + iy^lP_l}) \eqn\C$$
where $t=2\pi R_6$ and $y^l=2\pi {G^{l6}\over G^{66}}$, {\it i.e.}
$ y^1=2\pi\alpha, y^i=2\pi(\alpha\beta^i
+\gamma^i), l=1,..5,\, i=2,..5.$ 

One can also check that the expression  $-t{\cal H} +
iy^lP_l$ in  \C\  is exactly the
Hamiltonian derivable  from the (not fully 6d covariant)
Lagrangian that is  shown  in  [\Perry, \Schwarz] to give 
rise to the self-duality equation \A.

The strategy of the calculation is the following.
The partition function \C\ is by construction $SL(5,{\cal Z})$
invariant, due to the underlying $SO(5)$ invariance
in the coordinate space we have labelled $l=1...5$. We shall prove it is
also $SL(2,{\cal Z})$ invariant in the directions 1 and 6. The combination of
these two invariances will ensure $SL(6,{\cal Z})$ invariance.

As in any calculation of partition functions in string theory, we
divide the calculation of the trace into two parts. First we 
sum over the zero modes in \C\ and compute their anomaly under
$SL(2,{\cal Z})$, which will cancel against a similar anomaly in the  
sum over oscillators. A subtlety in performing this trace is to 
regularize  the vacuum energy in a way that
preserves the $SL(5,{\cal Z})$ invariance. 
The regularization is essential
for the  $SL(2,{\cal Z})$ anomaly cancellation. 
As we will see,
the latter is due only
to the contribution of the `2d-massless' modes, {\it i.e} the modes
with zero momentum in the transverse directions $2,\ldots ,5$. 
The modes with non-zero transverse momentum have a partition function of
2d massive bosons, which is $SL(2,{\cal Z})$ invariant by itself.

\chapter  {Transformation of the  Zero Modes of the Partition Function}

The trace in the partition function  
$$ Z(R_1, R_6, g_{ij}, \alpha,\beta^i, \gamma^i) = 
tr  \exp\{-t{\cal H} + i2\pi\alpha P_1 + i2\pi (\alpha\beta^i +
\gamma^i)P_i\} \eqn\trace$$
is over all independent Fock space operators which appear in
the normal mode expansion of the free massless tensor gauge field
$B_{MN}$. In this section we trace only on the zero mode operators and
determine their transformation under $SL(2,{\cal Z})$.
To this end,
we express the Hamiltonian \Ham\ and momentum \Pel\
in terms of the metric parameters in \B.
Neglecting temporarily the integration, we get
$$\eqalign{-t{\cal H} =& -{\pi \over 6} R_6  \sqrt{G_5}{G_5}^{ll'}
{G_5}^{mm'}{G_5}^{nn'} H_{lmn}H_{lmn}\cr
=&-{\pi \over 6} R_6R_1{\sqrt g}\,\bigl [(g^{ii'}g^{jj'}g^{kk'} +
{3\over R_1^2}\beta^i\beta^{i'}g^{jj'}g^{kk'})H_{ijk}H_{i'j'k'} 
\cr & \hskip5pt +{6\over R_1^2}\beta^{i'}g^{jj'}g^{kk'}H_{1jk}H_{i'j'k'} +
{3\over 2R_1^2} (g^{jj'}g^{kk'} - g^{jk'}g^{kj'})
H_{1jk}H_{1j'k'}\bigr ]\cr}\eqn\ham $$
and
$$P_1= -{1\over 8}\epsilon^{jkj'k'}H_{1jk}H_{1j'k'}\,;\qquad
P_i = -{1\over 4}\epsilon^{jkj'k'}H_{1jk}H_{ij'k'}\,.\eqn\pel$$
The terms in the exponential of \trace\ are then
$$\eqalign{-t{\cal H }+& i2\pi\alpha P_1 + i2\pi (\alpha\beta^i +
\gamma^i)P_i \cr
=&-{\pi \over 6}  R_6R_1{\sqrt g}g^{ii'}g^{jj'}g^{kk'}
H_{ijk}H_{i'j'K'} \cr &-{\pi\over 4} [{R_6\over R_1}{\sqrt g}(g^{jj'}g^{kk'} 
- g^{jk'}g^{kj'}) + i\alpha \epsilon^{jkj'k'}]H_{1jk}H_{1j'k'}
\cr &-{\pi\over 2}[{R_6\over R_1}{\sqrt g}(g^{jj'}g^{kk'} -
g^{jk'}g^{kj'}) + i\alpha \epsilon^{jkj'k'}]
\beta^i H_{1jk}H_{ij'k'} 
\cr &-{\pi\over 4}[{R_6\over R_1}{\sqrt g}(g^{jj'}g^{kk'} -
g^{jk'}g^{kj'}) + i\alpha \epsilon^{jkj'k'}] \beta^i
\beta^{i'}H_{ijk}H_{i'j'k'}
-{i\pi \over 2} \gamma^i \epsilon^{jkj'k'} H_{1jk} H_{ij'k'}\cr}\,.\eqn\hham$$
Notice that in the last line of \hham, the term 
$ \epsilon^{jkj'k'} \beta^i\beta^{i'}H_{ijk}H_{i'j'k'}$ is identically
zero. We retain it in order to write \hham\ in terms of
the matrix 
$${\cal A}^{jkj'k'} = {R_6\over R_1}{\sqrt g}(g^{jj'}g^{kk'} -
g^{jk'}g^{kj'}) + i\alpha \epsilon^{jkj'k'}.\eqn\bene$$
To sum over the zero modes, 
it is convenient to rewrite \hham\ as
$$\eqalign{&-t{\cal H }+ i2\pi\alpha P_1 + i2\pi (\alpha\beta^i +
\gamma^i)P_i
= \bigl ( -{\pi \over 6}  R_6R_1{\sqrt  g}g^{ii'}g^{jj'}g^{kk'}
H_{ijk}H_{i'j'k'}\cr 
&-{\pi\over 4 {|\tau |}^2}{R_6 \over
R_1}{\sqrt g} (g^{jj'}g^{kk'} - g^{jk'}g^{j'k})H_{ijk}H_{i'j'k'}
\gamma^i\gamma^{i'} 
-\pi {\cal A}^{jkj'k'} {1\over 4}[H_{1jk} + x_{jk}] 
[H_{1j'k'} + x_{j'k'}]\bigr ) \cr }\eqn\define$$
where
$$ x_{jk} \equiv \beta^i H_{ijk} + {i \over 4} \gamma^i
{{\cal A}^{-1}}_{jkj'k'}  \epsilon^{j'k'gh}H_{igh} 
\eqn\some$$ 
and 
$$\tau = \alpha + i {R_6 \over R_1}\, .\eqn\short$$

The form \define\ is particularly useful for the zero mode calculation 
since it  allows us to use a generalization of the 
Poisson summation formula [\Green]
$$\sum_{n \in {\cal Z}^p} 
e^{-\pi (n+x)\cdot A\cdot (n+x))} = (\det A)^{-\half}
\sum_{n \in {\cal Z}^p} e^{-\pi n\cdot A^{-1}\cdot n} e^{2\pi in\cdot x}\,.
\eqn\Poisson $$
The six fields $H_{1jk}$ have zero mode eigenvalues 
$n_1,\ldots , n_6\in {\cal Z}^6$, which we sum over 
in the Poisson formula.
The zero modes of the four fields $H_{ijk}$ are labelled by the integers
$n_7,\ldots , n_{10}$.

The partition function \trace\  can be written as 
$$Z= Z_{\rm osc} \cdot Z_{\rm zero\, modes}\eqn\zero$$ where
$Z_{\rm osc}$ is the contribution from the sum over the oscillators
which we compute in the next sections. The contribution from
the zero modes is 
$$\eqalign{Z_{\rm zero\, modes}  =&  
\sum_{n_7,\ldots , n_{10}\in {\cal Z}^4}
\exp\{-\pi { R_6R_1 \over 6}{\sqrt  g}g^{ii'}g^{jj'}g^{kk'}
H_{ijk}H_{i'j'k'}\cr & \hskip90pt -{\pi\over 4 {|\tau |}^2}{R_6 \over
R_1}{\sqrt g} (g^{jj'}g^{kk'} - g^{jk'}g^{j'k}) H_{ijk}H_{i'j'k'}
\gamma^i \gamma^{i'}\}\cr & \hskip30pt
\cdot\sum_{n_1,\ldots , n_6\in {\cal Z}^6} 
e^{-\pi (n+x)\cdot A\cdot (n+x)} \cr}
\eqn\zeroone $$ 
where $A_{11} = {\cal A}^{23 23}, \,A_{16}= {\cal A}^{23 45}, \ldots$, 
$x_1= x_{23}, \, x_2= x_{24},\ldots $ and 
$H_{123}=n_1$,
$H_{124}=n_2$, $H_{125}=n_3$, $H_{134}=n_4$, $H_{135}=n_5$, 
$H_{145}=n_6$. 
Using \Poisson\ we obtain
$$\eqalign{ Z=& Z_{osc} \cdot \sum_{n_7,\ldots , n_{10}\in {\cal Z}^4}
\exp\{-\pi { R_6R_1 \over 6}{\sqrt  g} g^{ii'}g^{jj'}g^{kk'}
H_{ijk}H_{i'j'k'}\cr & \hskip90pt -{\pi\over 4 {|\tau |}^2} {R_6 \over
R_1}{\sqrt g} (g^{jj'}g^{kk'} - g^{j'k}g^{jk'})H_{ijk}H_{i'j'k'}
\gamma^i\gamma^{i'}\} \cr 
& \hskip30pt \cdot\sum_{n_1,\ldots , n_6\in {\cal Z}^6}
\exp \{-{\pi\over 4 
{|\tau |}^2}[{R_6\over R_1}{\sqrt g}(g_{jj'}g_{kk'} -
g_{jk'}g_{kj'}) - i{\alpha \over g} \epsilon_{jkj'k'}]H_1^{jk}H_1^{j'k'}\cr 
&\hskip70pt+ \pi [iH_1^{jk}(\beta^i H_{ijk} + {{\cal
A}^{-1}}_{jkj'k'}\epsilon^{j'k'gh}H_{igh}\gamma^i {i\over 4}]\}
\cdot {1 \over {\sqrt{\det A}}} \cr}\eqn\almost $$
where the $H_1^{jk}$ are defined to be the integers $H_{1jk}$,
and ${\cal A}^{-1}$ is given by
$${{\cal A}^{-1}}_{jkj'k'} =  |\tau|^{-2} 
\bigl [{R_6\over R_1}{1\over \sqrt g}(g_{jj'}g_{kk'} -
g_{jk'}g_{kj'}) - i{\alpha \over g} \epsilon_{jkj'k'}\bigr
]\, ,\eqn\inverse$$
and  $A$ defined below \zeroone\ has $\det A=|\tau|^6$.
The expression \almost\ is only a rewriting of \zero\ and therefore
is equivalent to \trace. 
The Poisson transformed version \almost\ is particularly useful to 
exhibit the invariance of \trace\ under modular transformations. 
(We remark that $Z_{\rm zero\, modes}$ 
can be expressed by the Riemann theta function 
$\vartheta \left[{\vec 0\atop {\vec 0}}\right] (\vec 0 , \Omega)$
defined as in [\Mumford],
where the 10x10 symmetric non-singular complex matrix $\Omega$ 
can be reconstructed from 
\ham, \pel\ .)

The generalization of the usual $\tau \rightarrow\ -\tau^{-1}$
modular transformation that 
leaves invariant the line element
\B\   if $d\theta^1\nobreak\rightarrow d\theta^6, \,
d\theta^6 \rightarrow -d\theta^1,\,
d\theta^i \rightarrow d\theta^i,$ is
$$R_1\rightarrow R_1|\tau|,\, R_6  \rightarrow R_6|\tau|^{-1},\, \alpha
\rightarrow -{|\tau|}^{-2}\alpha, \,\beta^i\rightarrow \gamma^i,\,
\gamma^i \rightarrow -\beta^i ,\, g_{ij}\rightarrow g_{ij}\,. \eqn\modular$$
We do the transformation \modular\ on \hham\  
as follows: starting from \hham\ we write 
$$\eqalign{&{Z}(R_1|\tau|,{R_6 |\tau|^{-1}}, g_{ij},
-\alpha |\tau|^{-2},
\gamma^i, -\beta^i )\cr
&= Z'_{\rm osc}\cdot
\sum_{n_7...n_{10}} \exp\{-\pi { R_6R_1 \over 6}{\sqrt g} g^{ii'}g^{jj'}g^{kk'}
H_{ijk}H_{i'j'k'} \cr 
&\hskip90pt -{\pi\over 4|\tau|^2}{R_6\over R_1}{\sqrt g}(g^{jj'}g^{kk'} -
g^{jk'}g^{kj'}) \gamma^i \gamma^{i'}H_{ijk}H_{i'j'k'}\} \cr 
&\hskip40pt \cdot
\sum_{n_1...n_6}\exp\{-{\pi \over |\tau|^2} \bigl [{R_6\over R_1}{\sqrt g}
(g^{jj'}g^{kk'} -
g^{jk'}g^{kj'}) - i\alpha \epsilon^{jkj'k'}\bigr ]H_{1jk}H_{1j'k'}\cr
&\hskip90pt -{\pi\over 2|\tau|^2}\bigl [{R_6\over R_1}
{\sqrt g}(g^{jj'}g^{kk'} -
g^{jk'}g^{kj'}) - i\alpha \epsilon^{jkj'k'}\bigr ]
\gamma^i H_{1jk}H_{ij'k'}\cr 
&\hskip90pt
+{i\pi \over 2} \beta^iH_{1jk}\epsilon^{jkj'k'}H_{ij'k'}\}\cr}\eqn\transform$$
where $Z'_{\rm osc}$ is the transformed oscillator trace 
derived in sect.'s 4,5.
The first two lines of \transform\  and \almost\ are the
same. The rest can also be identified if one rotates the
$H_1^{jk}$ variables, defining  new variables ${\tilde H_{1jk}}$ via 
$$H_1^{jk}= {1\over 2}\epsilon^{jkj'k'}{\tilde H_{1j'k'}}.\eqn\ROT$$
Thus we have shown  that under the transformation \modular\ 
$${Z}_{\rm zero\, modes}(R_1|\tau|,{R_6 |\tau|^{-1}}, g_{ij},
-\alpha |\tau|^2, \gamma^i, -\beta^i)
= (\det A )^{{\scriptstyle{1\over 2}}}\,
{Z}_{\rm zero\, modes}(R_1, R_6, g_{ij}, \alpha, \beta^i, \gamma^i)\,.
\eqn\si$$
\chapter{The Oscillator Trace}

In this section we set up the trace calculation in detail and  find the
normal mode expansion of the `Hamiltonian' $-t{\cal H} + i2\pi\alpha
P_1 + i2\pi (\alpha\beta^i + \gamma^i)P_i$ appearing in \trace.
The result is given in (4.19) and (4.20).
It is convenient to introduce  
the dependent field strength $H_{6mn}$ given in \DEP\   and write
$$\eqalign{ -t{\cal H} + i2\pi\alpha
P_1 + i2\pi (\alpha\beta^i + \gamma^i)P_i  = &  
{i\pi\over 12}\int_0^{2\pi} d^5\theta H_{lrs}
\epsilon^{lrsmn} H_{6mn}\cr
=& {i\pi\over 2}\int_0^{2\pi} d^5\theta \sqrt{-G} 
H^{6mn} H_{6mn}\cr}\eqn\SUB$$
where
$H^{6mn} = {1\over 2\sqrt{-G}}\epsilon^{mnlrs} H_{lrs}$ from \A .

Let's define  $\Pi^{mn}(\vec\theta,\theta^6)$,
the field conjugate to  $B_{mn}(\vec\theta,\theta^6)$, starting 
from the 6d Lagrangian
for a general (non-self-dual) two-form 
$I_6=\int d^6\theta (-{\sqrt{-G}\over 24})H_{LMN} H^{LMN}$ 
$$\Pi^{mn}\nobreak={\textstyle{\delta I_6\over \delta \partial_6
B_{mn}}}  = -{\sqrt{-G}\over 4} H^{6mn}\,.\eqn\no$$ 
In terms of $\Pi^{mn}$  \SUB\ can be written as 
$${-i\pi \int_0^{2\pi} d^5\theta 
(\Pi^{mn} H_{6mn} + H_{6mn} \Pi^{mn})}\,,\eqn\OR$$
where we have chosen a specific field ordering. 
The commutation relations of the two-form 
and its conjugate field $\Pi^{mn}(\vec\theta,\theta^6)$ are assumed
to be the standard ones 
$$\eqalign{[\Pi^{rs}(\vec\theta,\theta^6),
B_{mn}(\vec\theta',\theta^6)]
=& -i\delta^5 (\vec\theta - \vec\theta') 
(\delta^r_m\delta^s_n - \delta^r_n\delta^s_m)\cr
[\Pi^{rs}(\vec\theta,\theta^6),
\Pi^{mn}(\vec\theta',\theta^6)]
=& [B_{rs}(\vec\theta,\theta^6),
B_{mn}(\vec\theta',\theta^6)] =0\cr}\eqn\COM$$

{}From the Bianchi identity $\partial_{[L}H_{MNP]} = 0$ and the fact
that \A\ implies $\partial^L H_{LMN} =0$, it follows that
a solution to \A\ is given by a solution to the homogeneous equations

$$\partial^L\partial_L B_{MN} = 0\,;\qquad \partial^L B_{LN} =0\,.\eqn\BOX$$ 

These have a plane wave solution

$$B_{MN}(\vec\theta,\theta^6) 
=f_{MN}(p) e^{ip\cdot\theta} + (f_{MN}(p) e^{ip\cdot\theta})^\ast\eqn\PW$$

when 
$$G^{LN}p_L p_N =0\,;\qquad p^L f_{LN} =0\,.\eqn\PPW$$

Using the metric on the six-torus \B\  and solving for $p_6$ from
\PPW, we get 
$$\eqalign{ p_6 =& -{G^{6m}\over G^{66}}p_m -i {\sqrt{{G_5^{mn}\over
G^{66}}p_m p_n}}
\cr = &  -\alpha p_1 - (\alpha\beta^i + \gamma^i) p_i
-i R_6{\sqrt{G_5^{mn}p_m p_n}} \cr}\eqn\OMEGA$$
where $2\le i\le \,; 1\le m,n\le 5$. 
Now consider a wave in the 1-direction with wave vector $p_i=0$,
$p_6 = (-\alpha-i{R_6\over R_1}) p_1$, 
and use the gauge invariance of $B_{MN}$, {\it i.e.}
$f_{MN}\rightarrow f'_{MN} = f_{MN} + p_M g_N - p_N g_M$ to fix
$f'_{6n}=0$. This   gauge choice
$$B_{6n} = 0\,,\eqn\GAUGE$$ is consistent with the
commutation relations \COM\ and reduces the number of components of
$B_{MN}$ from 15 to 10. 
 
Furthermore, the $\rho\sigma j$ component of \A\ , where
$\rho,\sigma = 6,1$  and $2\le j\le 5$, can be used to 
eliminate $f_{1j}$ in terms of the six $f_{ij}$ as
$$f_{1j} = {\textstyle{(\alpha + i{R_6\over R_1}\beta^k + \gamma^k})\over
(\alpha + i{R_6\over R_1})} f_{jk} = {\textstyle p^k f_{jk}\over p^1}\,.
\eqn \FIX$$
This satisfies \PPW\ .  Finally 
the $\rho j k$ component of \A\ expresses three of the $f_{ij}$ 
in terms of the remaining three, leaving just three independent polarization
tensors corresponding to the physical degrees of freedom
of the 6d chiral two form with Spin(4) content (3,1).

We can now  expand the free quantum tensor gauge field as
$$B_{mn} (\vec\theta, \theta^6) = {\rm zero\, modes} +
\sum_{\vec p\ne 0} ( f_{mn}^\kappa b_{\vec p}^\kappa e^{ip\cdot\theta}
+ f_{mn}^{\kappa\ast} b_{\vec p}^{\kappa\dagger} e^{-ip^\ast\cdot\theta})
\eqn\NME$$ where $1\le\kappa\le 3$, $1\le m,n\le 5$, 
$p_6$ is defined by \OMEGA\ , and the sum in \NME\ is on the dual
lattice $\vec p = p_m \in {\cal Z}^5\ne \vec 0$.
Since oscillators with different polarizations commute, we can treat
each polarization separately and cube the end result. 
Also, having already computed the zero mode contribution in sect.3, we will
drop the `zero mode' term here. Thus $$B_{mn}(\vec\theta, \theta^6) = 
\sum_{\vec p\ne 0} ( b_{\vec p nm}\,e^{ip\cdot\theta}
+ b_{\vec p mn}^{\dagger} e^{-ip^\ast\cdot\theta})\,,
\eqn\SNM$$ where $b_{\vec p nm}= f_{mn}^1 b_{\vec p}^1 $, $\,$
for example. From
the mode expansion for $B_{mn}$ in \SNM, 
since $H^{6mn} = {1\over 2\sqrt{-G}}\epsilon^{mnlrs} H_{lrs}$,
we can also write one for $H^{6mn}$ as 
$$\Pi^{mn} (\vec\theta, \theta^6)  = -{\sqrt{-G}\over4} H^{6mn} =
\sum_{\vec p\ne 0} ( c^{6mn}_{\vec p}\,e^{ip^\ast\cdot\theta}
+ c^{6mn\dagger}_{\vec p}\, e^{-ip\cdot\theta})\,.
\eqn\HNM$$
Then substituting \SNM\ and \HNM\ in the first term of \OR\ we find
$$\eqalign{-i\pi \int_0^{2\pi} d^5\theta \Pi^{mn} H_{6mn}
=& -i\pi (2\pi)^5
\sum_{\vec p \ne 0} ip_6 (c_{-\vec p}^{6mn} + c_{\vec p}^{6mn\dagger})
(b_{\vec p mn} + b_{-\vec p mn}^{\dagger})\cr}\,.\eqn\MOM$$ 
We compute the commutators of these oscillator combinations
$$\eqalign{\int {d^5\theta\over(2\pi)^5}\, e^{-ip_l\theta^l} 
B_{mn}(\vec\theta, 0) 
=& \, b_{\vec p \,mn} + b_{-\vec p \,mn}^\dagger\equiv
B_{\vec p \,mn}\cr
\int {d^5\theta\over(2\pi)^5} \,e^{ip_l\theta^l}
\Pi^{mn} (\vec\theta, 0) 
=& \, c_{-\vec p}^{6mn} + c_{\vec p}^{6mn\dagger}\equiv
{\cal C}_{\vec p}^{6mn\dagger}\cr}\eqn\OSC$$
so that 
$$\eqalign{[{\cal C}_{\vec p}^{6rs\dagger}, B_{\vec {p'} \,mn}] 
=& {1\over (2\pi)^{10}} \,\int_0^{2\pi} d^5\theta d^5\theta' 
e^{i\vec p\cdot\vec\theta - i \vec{p'}\cdot\vec\theta'} 
\,[\Pi^{rs} (\vec\theta, 0), B_{mn} (\vec\theta', 0)]\cr
=& -{i\over (2\pi)^5} \delta_{\vec p, \vec p'}
(\delta_m^r\delta_n^s - \delta_n^r\delta_m^s)\,.\cr}\eqn\OSCOM$$
Normalizing ${\cal C} B \equiv {i\over{(2\pi)^5}}\tilde{\cal C} \tilde B$,
so that
$$\eqalign{[\tilde{\cal C}_{\vec p}^{6rs\dagger}, \tilde B_{\vec p' \,mn}]
=& -\delta_{\vec p, \vec p'}
(\delta_m^r\delta_n^s - \delta_n^r\delta_m^s)\cr}\eqn\NOSCOM$$
we get
$$\eqalign{-i\pi \int_0^{2\pi} d^5\theta (\Pi^{mn} H_{6mn} +
H_{6mn} \Pi^{mn} )
=& -i\pi 
\sum_{\vec p \ne 0} p_6 (\tilde{\cal C}_{\vec p}^{6mn\dagger} 
\tilde B_{\vec p mn} +
\tilde B_{\vec p mn} \tilde{\cal C}_{\vec p}^{6mn\dagger} )\,.\cr}\eqn\TOR$$
Reinserting the polarization tensors normalized as 
$f^{\kappa rs}(p) f_{rs}^\lambda (p) = \delta^{\kappa\lambda}$
and using \NOSCOM\ , we find that \TOR\ becomes
$$\eqalign{&-2 i \pi\sum_{\vec p \ne 0} p_6 
{\cal C}_{\vec p}^{\kappa\dagger} B_{\vec p}^\lambda
f^{\kappa mn}(p) f^\lambda_{mn}(p)
- i\pi \sum_{\vec p \ne 0} p_6 f^{\kappa mn}(p) f^\kappa_{mn}(p)\cr
&=  -2 i \pi\sum_{\vec p \ne 0} p_6 
{\cal C}_{\vec p}^{\kappa\dagger} B_{\vec p}^\kappa
-i \pi \sum_{\vec p \ne 0} p_6 \delta^{\kappa\kappa}\cr
&=  -2i\pi \sum_{\vec p \ne 0} p_6 
{\cal C}_{\vec p}^{\kappa\dagger} B_{\vec p}^\kappa
-\pi R_6 \sum_{\vec p} \sqrt{G_5^{mn} p_m p_n}\,\,
\delta^{\kappa\kappa}\cr}
\eqn\ORVE$$
where $1\le\kappa,\lambda\le 3$ and 
$$[{\cal C}_{\vec p}^{\kappa\dagger}, B_{\vec p'}^\lambda] =
\delta^{\kappa\lambda}\,\delta_{\vec p,\vec p'}\eqn\AOSC$$
The ordering chosen in \OR\ gives rise to the vacuum energy as the 
second term in \ORVE\ . This term is necessary for modular invariance and
requires an $SL(5,{\cal{Z}})$ invariant regularization which is
derived in Appendix A.

\chapter {The Anomaly Cancellation}

{}From \ORVE\ and \trace\ the partition function is 
$$Z = Z_{\rm zero\, modes}\,\cdot\tr e^{-2i\pi \sum_{\vec p \ne 0} p_6
{\cal C}_{\vec p}^{\kappa\dagger} B_{\vec p}^\kappa
-\pi R_6 \sum_{\vec p } \sqrt{G_5^{mn} p_m p_n}\,\,
\delta^{\kappa\kappa}}\eqn\AGAIN $$
with $p_6$ given in \OMEGA\ and $p_l = n_l\in {\cal Z}^5$ due to the torus.
We can now use the standard Fock space argument
$$tr\omega^{\sum_p p a^\dagger_p a_p} 
=\prod_p\sum_{k=o}^\infty \langle k |\omega^{p a^\dagger_p a_p} | k\rangle
=\prod_p {\textstyle 1\over {1 - \omega^p}}\eqn\OSCTR$$
 to do the trace on the oscillators in \AGAIN. The answer is
$$\eqalign{Z =& Z_{\rm zero\,modes}\cdot \bigl ( e^{-\pi R_6 \sum_{\vec n} 
\sqrt{G_5^{lm} n_l n_m}}\, \prod_{\vec n\ne \vec 0}
{\textstyle 1\over{1- e^{-i2\pi p_6}}}\bigr )^3\,.\cr}\eqn\OPF$$
\OPF\ is manifestly $SL(5,{\cal Z})$ invariant since $p_6$ is.
(In $p_6$, $\,$ $G^{6m}$ is a contravariant 5-vector defined in \Gsixinv\ .)
However, the vacuum energy $\sqrt{G_5^{lm} n_l n_m}$ is a divergent
sum. In (A.5) in Appendix A we derive its $SL(5,Z)$
invariant regularization to obtain 

$$\eqalign{Z =& Z_{\rm zero\,modes}\cdot
\bigl ( e^{ R_6 \pi^{-3} \sum_{\vec n\ne \vec 0} {\sqrt{G_5}\over
(G_{lm}n^ln^m)^3}}\,
\prod_{\vec n\ne \vec 0}
{\textstyle 1\over{1- e^{-2\pi R_6 \sqrt{G_5^{lm}n_l n_m} + i 2\pi\alpha n_1
+ i 2\pi (\alpha\beta^i +\gamma^i)n_i}}}\bigr )^3\cr}\,\eqn\SLFIVE$$

\noindent where the sum on $\vec n$ is on the original lattice
$\vec n = n^l \in {\cal Z}^5\ne\vec 0$ and the product on $\vec n$ is on
the dual lattice $\vec n = n_l\in {\cal Z}^5\ne \vec 0$.
$Z_{\rm zero\, modes}$ is given in \zeroone\ . 
To understand how the $SL(2,{\cal Z})$ invariance of $Z$ works,
we separate the product on
$\vec n = (n,n_{\perp})\ne \vec 0$ into a product
on (all $n$, but $n_\perp\ne\nobreak (0,0,0,0)$)  
and on ($n\ne 0$, $n_{\perp} = (0,0,0,0))$, where
$n_\perp\equiv n_i$ and $n\equiv n_1$. Then 
\SLFIVE\ becomes  $$\eqalign{Z =& Z_{\rm zero\,modes}
\cdot \bigl ( e^{\textstyle{R_6\over\pi R_1}\zeta (2)} 
\prod_{n_1\ne 0}
{\textstyle 1\over{1- e^{2\pi i (\alpha n_1 + i {\textstyle {R_6\over R_1}} 
|n_1|)}}}\bigr )^3
\cr &\cdot
\bigl ( \prod_{n_i\ne (0,0,0,0)} e^{-2\pi R_6 <H>_{n_\perp}}
\, \prod_{n_1\in {\cal Z}}
{\textstyle 1\over{1- e^{-2\pi R_6 \sqrt{G_5^{lm}n_l n_m} + i 2\pi\alpha n_1
+ i 2\pi (\alpha\beta^i +\gamma^i)n_i}}}\bigr )^3\cr
=& Z_{\rm zero\,modes}\cdot 
\bigl (\eta (\tau) \bar\eta(\bar\tau)\bigr )^{-3}\cr
&\cdot \bigl ( \prod_{n_i\ne (0,0,0,0)} 
e^{-2\pi R_6 <H>_{n_\perp}} 
\prod_{n_1\in {\cal Z}}
{\textstyle 1\over{1- e^{-2\pi R_6 \sqrt{G_5^{lm}n_l n_m} + i 2\pi\alpha n_1
+ i 2\pi (\alpha\beta^i +\gamma^i)n_i}}}\bigr )^3\cr}\eqn\SLTWO$$
\noindent where $\tau\equiv \alpha + i{\textstyle {R_6\over R_1}}$ and
$<H>_{n_\perp}$ is given in (A.11).

\noindent In \SLTWO\ we have separated the contribution of the `2d
massless' scalars from the contribution of the
`2d massive' scalars. The former are the modes with zero momentum
$n_{\perp}=0$ in
the transverse direction $i=2...5$, which appear  as massless bosons
on the 2-torus in the directions $1$ and $6$.  
Instead, the modes associated with $n_{\perp} \ne 0$ correspond to
massive bosons on the 2-torus. Their partition function at fixed $n_{\perp}$ 
is $$e^{-2\pi R_6 <H>_{n_\perp}} \prod_{n_1\in {\cal Z}}
{\textstyle 1\over{1- e^{-2\pi R_6 \sqrt{G_5^{lm}n_l n_m} + i 2\pi\alpha n_1
+ i 2\pi (\alpha\beta^i +\gamma^i)n_i}}}\eqn\MB$$
and is $SL(2,{\cal Z})$ symmetric by itself, since there is no anomaly
for massive states.
We show in Appendix B  how
\MB\ can be derived from the path integral 
for a complex scalar field coupled to a constant gauge field
on the 2-torus.
The modular
invariance under \modular\ reduces on the 2-torus to the standard  
$\tau \rightarrow -{1\over \tau}$ transformation plus gauge invariance. 
In this path integral derivation the  
invariance of \MB\ under \modular\  then follows by construction.

The only piece of \SLTWO\ that has an
$SL(2,{\cal Z})$ anomaly is the one associated with the `2d
massless' modes 
$$e^{\textstyle{R_6\over\pi R_1}\zeta (2)}
\prod_{n_1\ne 0}
{\textstyle 1\over{1- e^{2\pi i (\alpha n_1 + i {\textstyle {R_6\over R_1}}
|n_1|)}}} = \bigl(\eta (\tau) \bar\eta(\bar\tau)\bigr )^{-1}$$
where the Dedekind eta function 
$\eta (\tau) \equiv 
e^{\pi i\tau\over{12}}\prod_{n=1}^\infty (1 - e^{2\pi i \tau n)}$,$\,$ 
and the Riemann zeta function $\zeta(2)={\textstyle{\pi^2\over 6}}$
results from Appendix A. 
Under the $SL(2,{\cal Z})$ transformation \modular\ of sect. 3,
$\tau\rightarrow -{1\over \tau}$ and  
$$(\eta (\tau) \bar\eta(\bar\tau))^{-3}\rightarrow 
|\tau|^{-3} (\eta (\tau) \bar\eta(\bar\tau))^{-3}\,. \eqn\cinque$$

This is how the oscillator anomaly cancels the zero mode anomaly
in \si. Hence the combination $Z_{\rm zero\, modes} \cdot
(\eta (\tau) \bar\eta(\bar\tau))^{-3}$ is $SL(2,{\cal Z})$ invariant. 
 
In analogy with the modular group $SL(2,{\cal Z})$ which can be generated by
two transformations such as $\tau\rightarrow \tau + 1$ and 
$\tau\rightarrow -{\textstyle 1\over\tau}$, the mapping class groups
of the $n$-torus, {\it i.e.}
the modular groups $SL(n,{\cal Z})$ can be generated
by just two transformations as well [\Coxeter].
In Appendix C we show how the 
$SL(5,{\cal Z})$ invariance of \SLFIVE\
and the $SL(2,{\cal Z})$ invariance of \SLTWO\ imply
symmetry under the $SL(6,{\cal Z})$ generators. 

\chapter{Conclusions}
In this paper we have computed explicitly the partition function of
the M-theory fivebrane chiral two-form on a six-torus.
In analogy with the  $SL(2,{\cal Z})$ modular invariance of 2d string theory, 
the fivebrane partition function  on $T^6$  has  $SL(6, {\cal
Z})$  invariance.  We have shown this in two
steps. We started with a manifestly $SL(5, {\cal Z})$  invariant
formalism (where 5 here refers to  the directions 1...5) 
and made sure that the regularization of the vacuum energy
did not spoil it. The crucial step, however, was to prove 
an additional $SL(2, {\cal Z})$ 
symmetry in the directions 1 and 6. This $SL(2, {\cal Z})$ invariance was
achieved by the cancellation of the anomaly from the 
sum over the zero modes with the anomaly of the 
`2d massless' modes from the oscillator trace. We then showed how
the combination of these symmetries implies $SL(6, {\cal Z})$ invariance. 
 
One of the problems with the M-theory fivebrane is that it is hard to
write down a manifestly covariant Lagrangian. Nonetheless, 
our result proves  that  the M-theory fivebrane chiral two-form 
can be consistently quantized on a six-torus.

Finally, we have shown that in this case
the partition function for the fivebrane two-form has no
dependence on the spin structure. Our result depends on the
fact that we compactify on $T^2\times T^4$ and would not
hold automatically on other spaces.

\vskip20pt
\leftline{\bf Acknowledgement}

We thank Edward Witten for valuable discussions, and
L.D. thanks the IAS for its hospitality. 
\vfill\eject

\Appendix{A}
\noindent $SL(5,Z)$ INVARIANT REGULARIZATION OF THE VACUUM ENERGY

The vacuum energy $<H>\equiv {1\over 2} \sum_{\vec p\in {\cal Z}^5}
\sqrt {{G_5}^{lm}
p_lp_m} = \half \sum_{\vec p\in {\cal Z}^5}|{\vec p}|$  appearing  in sect.3 is
a divergent sum and
needs to be regularized in an $SL(5,{\cal Z})$ invariant way. 
Here the sum is on the dual lattice $\vec p = p_l\in {\cal Z}^5$. 
The answer given in (A.5) is derived as follows. 
We rewrite $<H>$ as 
$$<H> = {1\over 2} \sum_{\vec p \in {\cal Z}^5}
{\sqrt {{G_5}^{lm}p_lp_m}} = {1\over 2}
\sum_{\vec p\in {\cal Z}^5}|\vec p| 
e^{i{\vec p} \cdot {\vec x}}\,\Big|_{{\vec x} =0}\,.
\eqn\aone $$ 
We  express $|\vec p|$ in terms of its 5d Fourier transform as 
$${|\vec p|} = \int d^5y e^{-i{\vec p}\cdot
{\vec y}} (-{2\over \pi^3} \sqrt{G_5}) \,{1\over |{\vec y}|^6}\,. \eqn\star$$
Then 
$$\eqalign{&\sum_{\vec p}|\vec p| e^{i{\vec p} \cdot {\vec x}}\cr &
= -{2\over \pi^3} \sqrt{G_5} \int d^5y  {1\over |{\vec y}|^6}
\sum_{\vec p}e^{i{\vec p}\cdot  ({\vec x} - {\vec y})} \cr
&= -{2\over \pi^3} \sqrt{G_5} \int d^5y  {1\over |{\vec y}|^6}
(2\pi)^5\sum_{{\vec n}\ne 0} \delta^5({\vec x}-{\vec y} +2\pi {\vec
n})\cr & = - 64 \pi^2  \sqrt{G_5} \sum_{{\vec n}\ne 0}
{1\over |{\vec x} + 2\pi{\vec n}|^6}\cr } \eqn\stop$$
where we have used the equality
$$\sum_{\vec p}e^{i{\vec p}\cdot {\vec x}} = (2\pi)^5
\sum_{\vec n} \delta^5({\vec x} +2\pi {\vec n}) \eqn\stim$$
and the sum on $\vec n$ is on the original lattice 
$\vec n = n^l\in {\cal Z}^5$.
Our regularization consists in removing the ${\vec n}=0$ term from this sum.
The regularized  vacuum energy  follows from \aone\ and \stop\
$$<H> = -{1\over 2\pi^4}{\sqrt G_5} \sum_{\vec n \ne 0} {1\over
(G_{lm}n^ln^m)^3} = -32\pi^2\sqrt{G_5}\sum_{{\vec n}\ne 0} {1\over
|2\pi {\vec n}|^6}\,. \eqn\still $$
We want to show now that in the
case of zero transverse momentum, {\it i.e.} $ n_{\perp} = (n_2, n_3, n_4, n_5)
=(0,0,0,0)$,  the regularization \still\
reduces to the usual $\zeta$ function regularization.
This  is essential to give the Dedekind eta
function in sect.5 for the anomaly cancellation with the
zero modes of sect.3.

To this purpose we rewrite the vacuum energy \still\ as a sum 
(on the dual lattice $p_\perp = p_{\perp i}\in {\cal Z}^4$) of terms
at fixed transverse momentum: 
$$<H> = -32\pi^2\sqrt{G_5}\sum_{p_{\perp}}{1\over (2\pi)^4}\int
d^4z_{\perp} e^{-ip_{\perp}\cdot 
z_{\perp}} \sum_{{\vec n}\ne 0} {1\over |2\pi\vec n + z_{\perp}|^6}\,.
\eqn\stew$$
The identity \stew\ can be checked from \still\ by using \stim.
By a change of variables $ z_{\perp} \rightarrow y_{\perp} =z_{\perp}
+ 2\pi {\vec n}$,  \stew\ can be now rewritten as
$$<H> = -32\pi^2\sqrt{G_5}\sum_{p_{\perp}}{1\over (2\pi)^4}\int d^4y_{\perp}
e^{-ip_{\perp}\cdot
y_{\perp}} \sum_{n^1\in{\cal Z}\ne 0} {1\over |2\pi n^1 + y_{\perp}|^6}
\eqn\newstew $$
where $|2\pi n^1 + y_{\perp}|^2\equiv 
[(2\pi n^1)^2 G_{11} + 2 (2\pi n^1) G_{1i} y_\perp^i + 
y_\perp^i y_\perp^j G_{ij}]$.
Now we consider the $p_{\perp} =0$ part only in \newstew\ and do  
the 4d integration in $d^4y_{\perp}$ with the result 
$$<H> = -32\pi^2\sqrt{G_5} {1\over (2\pi)^4} \sum_{n\in {\cal Z}}
{\pi^2\over g}{1\over
(2\pi n)^2}{1\over 2{R_1}^2}\,.\eqn\stall$$ 
Remembering that ${\sqrt{G_5}\over\sqrt{g}}= R_1$ and that
$\sum^\infty_{n=1} {1\over n^2}=
\zeta (2)={\pi^2 \over 6}$, we finally obtain 
$$<H>_{p_{\perp}=0} = -{1\over 12}{1\over R_1} = {1\over R_1}
\zeta (-1) .\eqn\stay$$
\vfill\eject

We note that computing the $d^4y_{\perp}$ integration in \newstew\ for
all $p_\perp$, we recover the spherical Bessel functions which appear
for massive bosons:
$$<H> = \sum_{p_\perp} <H>_{p_\perp}\eqn\PERP$$
where 
$$<H>_{p_\perp} = - |p_\perp |^2 R_1 \sum_{n^1=1}^\infty 
{\rm cos}(p_\perp\cdot\beta 2\pi n^1)
[ K_2(2\pi n^1 R_1 |p_\perp |) - K_0(2\pi n^1 R_1 |p_\perp |) ] \eqn\BES $$

For $\beta^i = 0$, \PERP\ can be expressed in terms of a standard
integral occurring in the effective potential for a 4d scalar field at
finite temperature:
$$<H> = - {1\over 2\pi^2 R_1} \sum_{p_\perp} \int_0^\infty dx 
{\textstyle x^2\over {\sqrt{x^2+a^2}}}
{\textstyle 1\over {(e^{\sqrt{x^2 + a^2}} -1})}\eqn\FITMP$$
where $a= 2\pi R_1 |p_\perp|$.

\vfill\eject

\Appendix{B}
\noindent{$SL(2,{\cal Z})$ SYMMETRIC PARTITION FUNCTION OF A 2D
MASSIVE SCALAR FIELD}

In (5.5)
we wrote the contribution from the oscillator trace to 
the  partition function of the chiral two-form on the torus as a
product over values of the transverse
momentum $n_\perp$. We show here that  
each term in the product with fixed $n_{\perp}\ne 0$ given in 
\MB\ 
is the square root of 
the partition function on $T^2$ of a massive complex scalar 
with $m^2 \equiv g^{ij} n_i n_j$ coupled to
a constant gauge field
$A^\mu \equiv i G^{\mu i}  n_i$ with $\mu,\nu ={1,6};\,i,j = 2,\ldots 5$. 
We show that \MB\
is a gauge transformation on the 2d gauge field 
combined with an $SL(2,{\cal Z})$ transformation on $T^2$.
The metric on $T^2$ is $h_{11} = R_1^2\, , h_{66} = R_6^2 +
\alpha^2 R_1^2\, , \, h_{16} = -\alpha R_1^2$. Its inverse is
$h^{11}= G^{11}$, $h^{66}=G^{66}$ and $h^{16}=G^{16}$. 
The invariance under \modular\
follows since the 2d partition function \MB\ can 
be derived from a manifestly $SL(2,{\cal Z})$ invariant {\it path integral}
on the 2-torus: 
We start with 
$$\eqalign{ {\rm P.I.}
&=\int d\phi \,d\bar\phi \,\,
e^{-\int_0^{2\pi} d\theta^1 \int_0^{2\pi} d\theta^6\,\,
h^{\mu\nu}(\partial_\mu + A_\mu )\bar\phi
(\partial_\nu - A_\nu) \phi + m^2 \bar\phi \phi}\cr
&=\int d\phi \,d\bar\phi \, 
e^{-\int_0^{2\pi} d\theta^1 \int_0^{2\pi} d\theta^6
\bar\phi ( -h^{\mu\nu}\partial_\mu
\partial_\nu + 2A^\mu\partial_\mu + G^{ij} n_i n_j ) \phi}\cr 
&=\int d\bar\phi\, d\phi 
e^{-\int_0^{2\pi} d\theta^1 \int_0^{2\pi} d\theta^6
\bar\phi (-(({1\over R^1})^2 + ({\alpha\over R_6})^2)\partial_1^2 
-({1\over R_6})^2\partial_6^2 -2 {\alpha\over {R_6^2}}
\partial_1\partial_6 + 2A^1\partial_1 + 2A^6\partial_6
+ G^{ij} n_i n_j )\phi}\cr  
& = \det \Bigl ( 
[- ( {\scriptstyle{1\over R_1^2}} + ({\scriptstyle {\alpha\over R_6}})^2 )
\partial_1^2 - ({\scriptstyle {1\over R_6}})^2\partial_6^2 
- 2\alpha ({\scriptstyle {1\over R_6}})^2\partial_1\partial_6  
+ G^{ij} n_i n_j  + 2 G^{1i}n_i \partial_1 
+ 2G^{6i}n_i \partial_6 ]\,\Bigr )^{-1}\cr
&= e^{- \tr \ln \Bigl [ 
- ( {\scriptstyle{1\over R_1^2}} + ({\scriptstyle {\alpha\over R_6}})^2 )
\partial_1^2 - ({\scriptstyle {1\over R_6}})^2\partial_6^2
- 2\alpha ({\scriptstyle {1\over R_6}})^2\partial_1\partial_6
+ G^{ij} n_i n_j  + 2 G^{1i}n_i \partial_1 
+ 2G^{6i}n_i \partial_6 \, \Bigr ] }\cr} \eqn\Buno$$

The trace on the momentum eigenfunctions of \Buno\ reduces to the sum
$$ {\rm P. I.} =
e^{-\sum_{s\in{\cal Z}}\sum_{r\in {\cal Z}} \Bigl [ \ln 
({4\pi^2 \over \beta^2}r^2 +  
( {\scriptstyle{1\over R_1^2}} + ({\scriptstyle {\alpha\over R_6}})^2 ) s^2
+ 2\alpha ({\scriptstyle {1\over R_6}})^2 r s 
+ G^{ij} n_i n_j  + 2 G^{1i}n_i s + 2G^{6i}n_i r )
\, \Bigr ]}
\eqn\btwo$$
where $\beta \equiv 2\pi R_6$. 
To evaluate the divergent sum on $r$,
we define $E^2 \equiv G^{lm}_5 n_l n_m$ where $n_1\equiv s$ and 
$$\eqalign{\nu (E) &= \sum_{r\in {\cal Z}} \ln({4\pi^2 \over \beta^2}r^2 
+ ( {\scriptstyle{1\over R_1^2}} + ({\scriptstyle {\alpha\over R_6}})^2 ) s^2
+ 2\alpha ({\scriptstyle {1\over R_6}})^2 r s
+ G^{ij} n_i n_j + 2 G^{1i}n_i s + 2G^{6i}n_i r)\cr
&= \sum_{r\in {\cal Z}} \ln \Bigl [{4\pi^2 \over \beta^2} 
(r + \alpha s + (\alpha\beta^i + \gamma^i) n_i)^2 
+ E^2 \Bigr ]\,.\cr}\,. 
\eqn\bthree$$
Then 
$$\eqalign{{\partial \nu (E)\over \partial E} & = 
\sum_r {2E\over {4\pi^2\over \beta^2 } (r + \alpha s + 
(\alpha\beta^i +\gamma^i) n_i )^2 
+ E^2} \cr
& = {\beta \sinh{\beta E}\over {\cosh{\beta E} - 
\cos{2\pi (\alpha s + (\alpha\beta^i +\gamma^i) n_i) }}}\cr
& = \partial_E\ln \Bigl [ \cosh{\beta E} - 
\cos{2\pi \bigl ( \alpha s + (\alpha\beta^i +\gamma^i) n_i \bigr)} \Bigr ]\cr}
\eqn\bfour$$
where we have used the fact that $\sum_{n\in {\cal Z}} 
{2y\over {(2\pi n + z)^2 +y^2}} = {\sinh{y}\over {\cosh y -\cos z}}$.
Integrating \bfour\ we get
$$\nu (E) = \ln \bigl [ \cosh{\beta E} - 
\cos{2\pi \bigl (\alpha s  + (\alpha\beta^i +\gamma^i) n_i\bigr )} \bigr ]
+ \ln 2 \eqn\bfive$$
where the integration constant in \bfive\
ensures an $SL(2,{\cal Z})$ invariant regularization of \btwo. 
Consequently for $n_1\equiv s$, we have that \Buno\ is 
$$\eqalign{({\rm P.I.})^{\half} &=  
\prod_{s\in {\cal Z}} {1\over 
\sqrt 2\sqrt {\cosh{\beta E} - 
\cos{2\pi ( \alpha s + (\alpha\beta^i +\gamma^i) n_i )}}}\cr
&= \prod_{s\in {\cal Z}} {e^{-{\beta E\over 2}}\over
{1 - e^{-\beta E + 2\pi i ( \alpha s + (\alpha\beta^i +\gamma^i) n_i )}}}\cr
&= e^{-\pi R_6 {\sum_{s\in {\cal Z}} \sqrt{G_5^{lm} n_l n_m}}}
\prod_{s\in {\cal Z}}
{1\over 
{1 - e^{-2\pi R_6 \sqrt{G_5^{lm} n_l n_m} 
+ 2\pi i  \alpha s 
+ 2\pi i (\alpha\beta^i + \gamma^i) n_i}}}\,.\cr
&= e^{-2\pi R_6 <H>_{n\perp}}
\prod_{n_1\in {\cal Z}}
{1\over
{1 - e^{-2\pi R_6 \sqrt{G_5^{lm} n_l n_m} 
+ 2\pi i\alpha n_1 
+ 2\pi i (\alpha\beta^i + \gamma^i) n_i}}}\,.\cr}
\eqn\bfinal$$
where $<H>_{n\perp}$ is a sum over spherical Bessel functions given in \BES\ . 
It is not at all obvious
that ${\rm P.I.}$ given by this formula is invariant under 
$\tau\rightarrow -\tau^{-1}$, where $\tau \equiv \alpha + i{R_6\over R_1}$
and $G_5^{lm}$ is given in \Gfiveinv\ , but it is true by construction.  
\vfill\eject
Furthermore \bfinal\ is invariant under the transformation \modular\ , which
also changes the 6d metric parameters $\beta^i$ and $\gamma^i$. 
Since $A_\mu\equiv h_{\mu\nu} i n_i G^{\nu i}$ where $\mu,\nu 
= {1,6}$, the transformation \modular\ on $A^\mu$ corresponds to
a gauge transformation $A^\mu\rightarrow A^\mu + \partial^\mu\lambda $,
and $ \phi \rightarrow e^{i\lambda}$, $\bar\phi\rightarrow e^{-i\lambda}$
where $$\eqalign{\lambda (\theta^1, \theta^6)
&= \theta^1 \Bigl [ {i\gamma^i\over {R_1^2 |\tau|^2}}
+ i {\alpha\over R_6^2} ({\alpha\gamma^i\over |\tau|^2} + \beta^i) \Bigr ]
- \theta^6 \, i {\alpha\gamma^i\over |\tau|^2}\cr}\,. \eqn\LAM $$
Hence \bfinal\ and thus \MB\ are invariant under \modular\ .

\vfill\eject

\Appendix{C}
\noindent GENERATORS OF $SL(n,{\cal Z})$

The $SL(n,{\cal Z})$ unimodular groups can each be generated by two
matrices[\Coxeter]. For $SL(6,{\cal Z})$ these can be chosen to be
$$U_1 = \left (\matrix{0&1&0&0&0&0\cr
0&0&1&0&0&0\cr
0&0&0&1&0&0\cr
0&0&0&0&1&0\cr
0&0&0&0&0&1\cr
1&0&0&0&0&0\cr}\right)\,;\qquad
U_2 = \left (\matrix{1&0&0&0&0&0\cr
1&1&0&0&0&0\cr
0&0&1&0&0&0\cr
0&0&0&1&0&0\cr
0&1&0&0&1&0\cr
0&0&0&0&0&1\cr}\right)\,.\eqn\U$$
That is every matrix $M$ in
$SL(n,{\cal Z})$ can be written 
as a product $U_1^{n_1} U_2^{n_2} U_1^{n_3}\dots$.
The matrices $U_1$ and $U_2$ act on the basis vectors of the
six-torus $\vec\alpha_I$ where $\vec\alpha_I\cdot\vec\alpha_J
= G_{IJ}$. For our metric \Gsix\ , the transformation
$$\left(\matrix{\vec\alpha'_1\cr
\vec\alpha'_2\cr
\vec\alpha'_3\cr
\vec\alpha'_4\cr
\vec\alpha'_5\cr
\vec\alpha'_6\cr}\right)
=U_2\left(\matrix{\vec\alpha_1\cr
\vec\alpha_2\cr
\vec\alpha_3\cr
\vec\alpha_4\cr
\vec\alpha_5\cr
\vec\alpha_6\cr}\right)\eqn\TPLUSONE$$
corresponds to
$$R_1\rightarrow R_1 , R_6  \rightarrow R_6, \alpha
\rightarrow \alpha - 1, \beta^i\rightarrow \beta^i,
\gamma^i \rightarrow \gamma^i+\beta^i, g_{ij}\rightarrow
g_{ij}\eqn\TPOMET$$
which leaves invariant the line element \B\
if $d\theta^1\nobreak\rightarrow d\theta^1 - d\theta^6, \,
d\theta^6 \rightarrow d\theta^6,\,
d\theta^i \rightarrow d\theta^i.$
$U_2$ is the generalization of the usual $\tau\rightarrow\tau - 1$
modular transformation.
It is easily checked that $U_2$ is an invariance of the partition function
\SLFIVE\ and \zeroone.
The less trivial generator $U_1$ can be related to the tranformation
\modular\ that we study in the text as follows:
$$U_1 = U' M_5\eqn\TONEOVERT$$
where $M_5$ is an $SL(5,{\cal Z})$ transformation given by
$$M_5 = \left (\matrix{0&0&-1&0&0&0\cr
0&1&0&0&0&0\cr
0&0&0&1&0&0\cr
0&0&0&0&1&0\cr
0&0&0&0&0&1\cr
1&0&0&0&0&0\cr}\right)\eqn\Mfive$$
and
$U'$ is the matrix corresponding to \modular\ :
$$U' = \left (\matrix{0&1&0&0&0&0\cr
-1&0&0&0&0&0\cr
0&0&1&0&0&0\cr
0&0&0&1&0&0\cr
0&0&0&0&1&0\cr
0&0&0&0&0&1\cr}\right)\,.\eqn\UMOD$$
Hence the $SL(5,{\cal Z})$ symmetry of \SLTWO\ together with its
invariance under the modular transformation \modular\ implies via \TONEOVERT\
invariance under the $SL(6,{\cal Z})$ generator $U_1$.
So due to its symmetry under both generators, the
partition function is invariant under the modular group
$SL(6,{\cal Z})$, the mapping class group of the six-torus.

\refout

\end

\input phyzzx
\overfullrule=0pt
\tolerance=5000
\twelvepoint
\def\tilde{\widetilde}

\pubnum{IASSNS-HEP-98/43}
\date{May 1998}
\titlepage
\title{A Modular Invariant Partition Function for the Fivebrane}
\vglue-.25in
\author{ Louise Dolan \foot{Research supported in part by the U.S. Department
of Energy under Grant No. DE-FG 05-85ER40219/Task A. 
Email: dolan@physics.unc.edu}} 
\address{Department of Physics and Astronomy
\break University of North Carolina
\break Chapel Hill
\break North Carolina 27599-3255}
\author{Chiara R. Nappi\foot{Research supported in part by
the Ambrose Monell Foundation. Email: nappi@ias.edu}}
\medskip
\address{School of Natural Sciences
\break Institute for Advanced Study
\break Olden Lane
\break Princeton, NJ 08540}
\bigskip

\parskip=6pt

\abstract{We compute an $SL(6,{\cal Z})$ invariant partition function
for the chiral two-form of the M theory fivebrane
compactified on the six-torus $T^6$. From a manifestly
$SL(5,{\cal Z})$ invariant formalism, we prove that the partition
function has  an additional
$SL(2,{\cal Z})$ symmetry. The combination of these two symmetries
ensures $SL(6,{\cal Z})$ invariance. Thus, whether or not a fully
covariant Lagrangian is  available, the fivebrane on the six-torus has a
consistent quantum theory.}

\REF\Witten{E. Witten, ``Five-brane Effective Action in M-theory,''
J. Geom. Phys. {\bf 22} (1997) 103; hep-th 9610234.}

\REF\Perry{M. Perry and J. H. Schwarz, ``Interacting Chiral Gauge
Fields in Six Dimensions and Born-Infeld Theory','' 
Nucl. Phys. {\bf B489} (1997) 47; hep-th/9611065.}

\REF\Schwarz{ John H. Schwarz, ``Coupling a Self-dual Tensor to
Gravity in Six Dimensions,''Phys. Lett. {\bf B395} (1997) 191; 
hep-th/ 9701008.}

\REF\PST{P. Pasti, D. Sorokin and M. Tonin, ``Covariant Action for a
D=11 Five-Brane with Chiral Field,'' Phys. Lett. {\bf B398} (1997) 41;
hep-th 9701037; ``On Lorentz
Invariant Actions for Chiral P-Forms,'' Phys.
Rev. {\bf D52} (1995) 4277; hep-th/9711100.}
 
\REF\Aganagic{ M. Aganagic, J. Park, C. Popescu and J. H. Schwarz,
``Worldvolume Action of the M-theory Fivebrane,'' Nucl. Phys. {\bf
B496} (1997) 191; hep-th/9701166.} 

\REF\Howe{ P.S. Howe, E. Sezgin and P.C. West, ``Covariant Field
Equations of the M-theory Five-brane, Nucl.Phys. {\bf B496} (1997)
191.}

\REF\Bandos{ I. Bandos, K. Lechner, A. Nurmagambetov, P. Pasti,
D. Sorokin and M. Tonin, ``On the Equivalence of Different
Formulations of M-theory Fivebrane,'' Phys. Lett. {\bf 408B} (1997)
135.}

\REF\Berg{E. Bergshoeff, D. Sorokin and P.K. Townsend, ``The M-5brane
Hamiltonian,'' hep-th/9805065.}   

\REF\hop {M. Hopkins and I. M. Singer, to appear}

\REF\Green{M.B. Green, J. H. Schwarz and E. Witten, 
$\underline{\rm Superstring\, Theory}$,
vol. I and II, Cambridge University Press: Cambridge, U.K. 1987.}

\REF\Coxeter{ H.S.M. Coxeter and W.O.J. Moser,
$\underline{\rm Generators\, and\, Relations\, for\, Discrete\, Groups}$,
Springer-Verlag: New York 1980.}

\chapter {Introduction}

The physical degrees of freedom of the 6d world volume theory of the M
theory fivebrane consist of an N=(2,0) tensor supermultiplet. This
multiplet contains a chiral two-form  
$B_{MN}$ with a self-dual three-form field strength 
$H_{LMN}= \partial_LB_{MN} + \partial_MB_{NL} + \partial_NB_{LM}$ 
where $1\le L,M,N \le 6$.
 
Many attempts
have been made recently to write a manifestly covariant
six-dimensional action for $H_{LMN}$ [\Perry 
--\Berg]. The difficulty is related to the self-duality
of the three-form field strength and is analogous to 
the  well-known problem of writing down covariant
actions for theories with chiral bosons. We adopt in this paper a
different approach. Rather than considering the Lagrangian, 
we ask if there is a modular invariant partition function. 

It is known there is no modular invariant partition function for a 
single chiral field in two dimensions. This is the reason, a posteriori,
that one cannot write a  covariant Lagrangian for such a field.
Such a Lagrangian, if it were to exist, could then be quantized
on a Riemann surface of genus $g$ and would yield results that depended only
on the metric of the Riemann surface in a modular invariant way.
Instead, it is known that a chiral scalar on a Riemann surface of
genus $g$ has $2^{2g}$ candidate partition functions.
The situation is expected to be similar for the partition function of
the M-theory fivebrane [\Witten].

Nonetheless in this paper  we compute 
the partition function $Z$ for a free self-dual $H_{LMN}$ field strength
on $T^6$ and show that  it is invariant 
under the $SL(6,{\cal Z})$ mapping class group of $T^6$. 
We circumvent the lack of a fully covariant Lagrangian by 
writing the self-dual three-form in six dimensions as  
an anti-symmetric three-form in five dimensions, which is no longer self-dual. 
In this formalism, the partition function is automatically $SL(5,{\cal Z})$
invariant. 
We show it has an additional
$SL(2,{\cal Z})$ invariance and from this prove that  
$Z$ is an example of an $SL(6,{\cal Z})$ automorphic form, 
even though only $SL(5,{\cal Z})$ symmetry is explicit.
It is given in (5.4) together with (3.11). 

The reason this partition function avoids the problems of the
chiral boson and manages to be modular invariant is that
we are compactifying on $T^2\times T^4$. From the point of view of $T^2$,
the three degrees of freedom of the two-form potential (which is the
(3,1) representation of the Spin(4)$\cong SU(2)\times SU(2)$ little
group for 6d, $N=(2,0)$ massless states)
behave like three 2d-massive scalars, therefore mimicking the situation  
of three non-chiral bosons.
This would not have happened had we compactified on $T^2\times {\bf CP2}$
for example. This is the way our  calculation and hence our result are
special to the compactification on $T^6$.

It was pointed out in the introduction to
[\Witten] that for a fivebrane on $\Sigma\times
{\bf CP}^2$ (with $\Sigma$ a Riemann surface) the chiral two-form
partition function depends on a spin structure on $\Sigma$. 
Our result shows
that in the case of $T^6$, there is no such spin structure dependence.
We understand
that, by relating the description of [\Witten] to the Kervaire
invariant,
M. Hopkins and I. M. Singer have determined when there is or is not a
dependence
of the chiral two-form partition function on the spin structure [\hop].
Notice  that 
for the partition function of a chiral two-form to depend on a spin
structure does not violate any symmetry of M-theory, since M-theory anyway
has fermions that require a spin structure. 
               
Although our result is restricted to the case of compactification on
the torus, we think it is interesting to exhibit an $SL(6, {\cal Z})$ 
modular invariant partition function for the M5-brane, {\it i.e.} a
quantum theory of the free 6d self-dual two-form with symmetry analogous to
the modular invariance of consistent interacting strings.  
In sect.3 we calculate the zero-mode contribution to $Z$, and in
sect.'s 4,5 we compute the oscillator trace and show how 
the $SL(2,{\cal Z})$  anomalies cancel. 

Appendix A describes an $SL(5, {\cal Z})$ invariant regularization
of  the vacuum energy in the trace over oscillators.
Appendix B reviews how
`2-d massive' part of our partition function is  $SL(2, {\cal Z})$ invariant,
an important ingredient of our proof.
Finally,  Appendix C deals with 
the generators of the mapping class groups
$SL(n,{\cal Z})$ for the n-torus and shows how $SL(2, {\cal Z})$ and 
$SL(5, {\cal Z})$ combine to give $SL(6, {\cal Z})$ invariance.

\chapter {Notation and Strategy}

The 6d chiral two-form $B_{MN}$ has a self-dual field-strength which satisfies
$$H_{LMN}(\vec\theta,\theta^6)= {1\over{6\sqrt{-G}}}G_{LL'}G_{MM'}G_{NN'}
\epsilon^{L'M'N'RST}H_{RST}(\vec\theta,\theta^6) \eqn\A $$
where $G_{MN}$ is the 6d metric.
With use of this equation of motion it is possible [\Schwarz] 
to eliminate the components $H_{6mn}$
in terms of the other components $H_{lmn}$ with $l,m,n
=1...5$ as follows
$$H_{6mn} = -{G^{6l}\over G^{66}} H_{lmn}
+ {1\over{6\sqrt{-G}G^{66}}}\epsilon_{mnlrs}H^{lrs}, \eqn\DEP $$
where now indices are raised with the 5d-metric $G_5^{mn}$ and 
$\epsilon_{12345} \equiv G_5\epsilon^{12345} = G_5$.
 
The Hamiltonian and momenta of the $H_{lmn}$ 
can be written in a fully 5d covariant way:
$${\cal H} = {1\over 12}{\int_0}^{2\pi}
d\theta^1...d\theta^5{\sqrt G_5}{G_5}^{ll'}
{G_5}^{mm'}{G_5}^{nn'} H_{lmn}(\vec\theta,\theta^6)\, 
H_{l'm'n'}(\vec\theta,\theta^6) \eqn\Ham$$
$$P_l = -{1\over 24}{\int_0}^{2\pi}
d\theta^1...d\theta^5 \epsilon^{rsumn} H_{umn}(\vec\theta,\theta^6)\,
H_{lrs}(\vec\theta,\theta^6)\eqn\Pel$$
with $1\le l,m,n,r,s,u\le 5$.

A general metric on $T^6$ is a function of 21 parameters and 
can be represented by the line element
$$\eqalign{ds^2 = & {R_1}^2(d\theta^1 -\alpha d\theta^6)^2 +
{R_6}^2(d\theta^6)^2 \cr
+&\sum_{i,j=2...5}g_{ij}(d\theta^i -
\beta^id\theta^1-\gamma^id\theta^6)
(d\theta^j - \beta^jd\theta^1 - \gamma^jd\theta^6) \cr}\eqn\B$$
where $0\le\theta^I\le 2\pi$, $1\le I\le 6$ and 
we single out  directions 1 and 6. The latter is our time direction.
The 21 parameters are as follows: $R_1$ and $R_6$ are the radii for 
directions 1 and 6,  
$g_{ij}$ is a 4d metric, $\beta^i,\gamma^j$ are the angles between
directions  1 and $i$, and between 6 and $j$, respectively; and  
$\alpha$ is related to the angle between 1 and 6, see (2.6).
The metric can be read off from the line element above to be  
$$\eqalign{ &{G}_{ij}= g_{ij}\,;
\qquad \hskip20pt G_{11}= {R_1}^2 + g_{ij}\beta^i\beta^j\cr 
&G_{i1}= -g_{ij}\beta^j\,;
\qquad G_{66}= {R_6}^2 + \alpha^2R_1^2 +g_{ij}\gamma^i\gamma^j\cr  
&G_{i6}= -g_{ij}\gamma^j\,;
\qquad G_{16}= -\alpha R_1^2 + g_{ij}\beta^i\gamma^j\,. \cr } \eqn\Gsix$$ 
Its inverse is 
$$\eqalign{G^{ij}=& {g}^{ij} +   {{\beta^i\beta^j}\over R_1^2}
+ {1  \over R_6^2}(\alpha \beta^i + \gamma^i)(\alpha \beta^j +
\gamma^j)\,;\hskip10pt
\qquad G^{11} = {1\over R_1^2}  + {\alpha^2 \over R_6^2}\cr
G^{1i}=& {\beta^i\over R_1^2} + {\alpha  \over R_6^2}(\alpha \beta^i +
\gamma^i)\,;
\qquad \hskip100pt G^{66} = {1\over R_6^2}\cr
G^{i6} =&{{(\alpha \beta^i + \gamma^i)} \over R_6^2}\,; 
\hskip150pt \qquad
G^{16} = {\alpha \over R_6^2}\,.\cr}\eqn\Gsixinv$$
The 5-dimensional inverse is  
$$\eqalign{ {G_5}^{ij}=& g^{ij} + {{\beta^i\beta^j}\over R_1^2}\,;\qquad
{G_5}^{i1}={\beta^i\over R_1^2}\,;\qquad
{G_5}^{11}= {1\over R_1^2}\cr} \eqn\Gfiveinv$$
where ${g}^{ij}$ is the 4d inverse of $g_{ij}$. Useful formulae are 
$G_5 = {R_1}^2 g, $ where $G_5 = {\rm det} G_{mn}$ 
and $g = {\rm det} g_{ij}$. Also $G=\det G_{IJ}$. 

We want to compute the  partition function on the twisted torus with metric
\Gsix.  As in string theory [\Green],  the partition function 
is given in terms of the Hamiltonian and momenta  by 
$$Z= tr(e^{-t{\cal H} + iy^lP_l}) \eqn\C$$
where $t=2\pi R^6$ and $y^l={G^{l6}\over G^{66}}$, {\it i.e.}
$ y^1=2\pi\alpha, y^i=2\pi(\alpha\beta^i
+\gamma^i), l=1,..5,\, i=2,..5.$ 

One can also check that the expression  $-t{\cal H} +
iy^lP_l$ in  \C\  is exactly the
Hamiltonian derivable  from the (not fully 6d covariant)
Lagrangian that is  shown  in  [\Perry, \Schwarz] to give 
rise to the self-duality equation \A.

The strategy of the calculation is the following.
The partition function \C\ is by construction $SL(5,{\cal Z})$
invariant, due to the underlying $SO(5)$ invariance
in the coordinate space we have labelled $l=1...5$. We shall prove it is
also $SL(2,{\cal Z})$ invariant in the directions 1 and 6. The combination of
these two invariances will ensure $SL(6,{\cal Z})$ invariance.

As in any calculation of partition functions in string theory, we
divide the calculation of the trace into two parts. First we 
sum over the zero modes in \C\ and compute their anomaly under
$SL(2,{\cal Z})$, which will cancel against a similar anomaly in the  
sum over oscillators. A subtlety in performing this trace is to 
regularize  the vacuum energy in a way that
preserves the $SL(5,{\cal Z})$ invariance. 
The regularization is essential
for the  $SL(2,{\cal Z})$ anomaly cancellation. 
As we will see,
the latter is due only
to the contribution of the `2d-massless' modes, {\it i.e} the modes
with zero momentum in the transverse directions $2,\ldots ,5$. 
The modes with non-zero transverse momentum have a partition function of
2d massive bosons, which is $SL(2,{\cal Z})$ invariant by itself.

\chapter  {Transformation of the  Zero Modes of the Partition Function}

The trace in the partition function  
$$ Z(R_1, R_6, g_{ij}, \alpha,\beta^i, \gamma^i) = 
tr  \exp\{-t{\cal H} + i2\pi\alpha P_1 + i2\pi (\alpha\beta^i +
\gamma^i)P_i\} \eqn\trace$$
is over all independent Fock space operators which appear in
the normal mode expansion of the free massless tensor gauge field
$B_{MN}$. In this section we trace only on the zero mode operators and
determine their transformation under $SL(2,{\cal Z})$.
To this end,
we express the Hamiltonian \Ham\ and momentum \Pel\
in terms of the metric parameters in \B.
Neglecting temporarily the integration, we get
$$\eqalign{-t{\cal H} =& -{\pi \over 6} R_6  \sqrt{G_5}{G_5}^{ll'}
{G_5}^{mm'}{G_5}^{nn'} H_{lmn}H_{lmn}\cr
=&-{\pi \over 6} R_6R_1{\sqrt g}\,\bigl [(g^{ii'}g^{jj'}g^{kk'} +
{3\over R_1^2}\beta^i\beta^{i'}g^{jj'}g^{kk'})H_{ijk}H_{i'j'k'} 
\cr & \hskip5pt +{6\over R_1^2}\beta^{i'}g^{jj'}g^{kk'}H_{1jk}H_{i'j'k'} +
{3\over 2R_1^2} (g^{jj'}g^{kk'} - g^{jk'}g^{kj'})
H_{1jk}H_{1j'k'}\bigr ]\cr}\eqn\ham $$
and
$$P_1= -{1\over 8}\epsilon^{jkj'k'}H_{1jk}H_{1j'k'}\,;\qquad
P_i = -{1\over 4}\epsilon^{jkj'k'}H_{1jk}H_{ij'k'}\,.\eqn\pel$$
The terms in the exponential of \trace\ are then
$$\eqalign{-t{\cal H }+& i2\pi\alpha P_1 + i2\pi (\alpha\beta^i +
\gamma^i)P_i \cr
=&-{\pi \over 6}  R_6R_1{\sqrt g}g^{ii'}g^{jj'}g^{kk'}
H_{ijk}H_{i'j'K'} \cr &-{\pi\over 4} [{R_6\over R_1}{\sqrt g}(g^{jj'}g^{kk'} 
- g^{jk'}g^{kj'}) + i\alpha \epsilon^{jkj'k'}]H_{1jk}H_{1j'k'}
\cr &-{\pi\over 2}[{R_6\over R_1}{\sqrt g}(g^{jj'}g^{kk'} -
g^{jk'}g^{kj'}) + i\alpha \epsilon^{jkj'k'}]
\beta^i H_{1jk}H_{ij'k'} 
\cr &-{\pi\over 4}[{R_6\over R_1}{\sqrt g}(g^{jj'}g^{kk'} -
g^{jk'}g^{kj'}) + i\alpha \epsilon^{jkj'k'}] \beta^i
\beta^{i'}H_{ijk}H_{i'j'k'}
-{i\pi \over 2} \gamma^i \epsilon^{jkj'k'} H_{1jk} H_{ij'k'}\cr}\,.\eqn\hham$$
Notice that in the last line of \hham, the term 
$ \epsilon^{jkj'k'} \beta^i\beta^{i'}H_{ijk}H_{i'j'k'}$ is identically
zero. We retain it in order to write \hham\ in terms of
the matrix 
$${\cal A}^{jkj'k'} = {R_6\over R_1}{\sqrt g}(g^{jj'}g^{kk'} -
g^{jk'}g^{kj'}) + i\alpha \epsilon^{jkj'k'}.\eqn\bene$$
To sum over the zero modes, 
it is convenient to rewrite \hham\ as
$$\eqalign{&-t{\cal H }+ i2\pi\alpha P_1 + i2\pi (\alpha\beta^i +
\gamma^i)P_i
= \bigl ( -{\pi \over 6}  R_6R_1{\sqrt  g}g^{ii'}g^{jj'}g^{kk'}
H_{ijk}H_{i'j'k'}\cr 
&-{\pi\over 4 {|\tau |}^2}{R_6 \over
R_1}{\sqrt g} (g^{jj'}g^{kk'} - g^{jk'}g^{j'k})H_{ijk}H_{i'j'k'}
\gamma^i\gamma^{i'} 
-\pi {\cal A}^{jkj'k'} {1\over 4}[H_{1jk} + x_{jk}] 
[H_{1j'k'} + x_{j'k'}]\bigr ) \cr }\eqn\define$$
where
$$ x_{jk} \equiv \beta^i H_{ijk} + {i \over 4} \gamma^i
{{\cal A}^{-1}}_{jkj'k'}  \epsilon^{j'k'gh}H_{igh} 
\eqn\some$$ 
and 
$$\tau = \alpha + i {R_6 \over R_1}\, .\eqn\short$$

The form \define\ is particularly useful for the zero mode calculation 
since it  allows us to use a generalization of the 
Poisson summation formula [\Green]
$$\sum_{n \in {\cal Z}^p} 
e^{-\pi (n+x)\cdot A\cdot (n+x))} = (\det A)^{-\half}
\sum_{n \in {\cal Z}^p} e^{-\pi n\cdot A^{-1}\cdot n} e^{2\pi in\cdot x}\,.
\eqn\Poisson $$
The six fields $H_{1jk}$ have zero mode eigenvalues 
$n_1,\ldots , n_6\in {\cal Z}^6$, which we sum over 
in the Poisson formula.
The zero modes of the four fields $H_{ijk}$ are labelled by the integers
$n_7,\ldots , n_{10}$.

The partition function \trace\  can be written as 
$$Z= Z_{\rm osc} \cdot Z_{\rm zero\, modes}\eqn\zero$$ where
$Z_{\rm osc}$ is the contribution fron the sum over the oscillators
which we compute in the next sections. The contribution from
the zero modes is 
$$\eqalign{Z_{\rm zero\, modes}  =&  
\sum_{n_7,\ldots , n_{10}\in {\cal Z}^4}
\exp\{-\pi { R_6R_1 \over 6}{\sqrt  g}g^{ii'}g^{jj'}g^{kk'}
H_{ijk}H_{i'j'k'}\cr & \hskip90pt -{\pi\over 4 {|\tau |}^2}{R_6 \over
R_1}{\sqrt g} (g^{jj'}g^{kk'} - g^{jk'}g^{j'k}) H_{ijk}H_{i'j'k'}
\gamma^i \gamma^{i'}\}\cr & \hskip30pt
\cdot\sum_{n_1,\ldots , n_6\in {\cal Z}^6} 
e^{-\pi (n+x)\cdot A\cdot (n+x)} \cr}
\eqn\zeroone $$ 
where $A_{11} = {\cal A}^{23 23}, \,A_{16}= {\cal A}^{23 45}, \ldots$, 
$x_1= x_{23}, \, x_2= x_{24},\ldots $ and 
$H_{123}=n_1$,
$H_{124}=n_2$, $H_{125}=n_3$, $H_{134}=n_4$, $H_{135}=n_5$, 
$H_{145}=n_6$. 
Using \Poisson\ we obtain
$$\eqalign{ Z=& Z_{osc} \cdot \sum_{n_7,\ldots , n_{10}\in {\cal Z}^4}
\exp\{-\pi { R_6R_1 \over 6}{\sqrt  g} g^{ii'}g^{jj'}g^{kk'}
H_{ijk}H_{i'j'k'}\cr & \hskip90pt -{\pi\over 4 {|\tau |}^2} {R_6 \over
R_1}{\sqrt g} (g^{jj'}g^{kk'} - g^{j'k}g^{jk'})H_{ijk}H_{i'j'k'}
\gamma^i\gamma^{i'}\} \cr 
& \hskip30pt \cdot\sum_{n_1,\ldots , n_6\in {\cal Z}^6}
\exp \{-{\pi\over 4 
{|\tau |}^2}[{R_6\over R_1}{\sqrt g}(g_{jj'}g_{kk'} -
g_{jk'}g_{kj'}) - i{\alpha \over g} \epsilon_{jkj'k'}]H_1^{jk}H_1^{j'k'}\cr 
&\hskip70pt+ \pi [iH_1^{jk}(\beta^i H_{ijk} + {{\cal
A}^{-1}}_{jkj'k'}\epsilon^{j'k'gh}H_{igh}\gamma^i {i\over 4}]\}
\cdot {1 \over {\sqrt{\det A}}} \cr}\eqn\almost $$
where the $H_1^{jk}$ are defined to be the integers $H_{1jk}$,
and ${\cal A}^{-1}$ is given by
$${{\cal A}^{-1}}_{jkj'k'} =  |\tau|^{-2} 
\bigl [{R_6\over R_1}{1\over \sqrt g}(g_{jj'}g_{kk'} -
g_{jk'}g_{kj'}) - i{\alpha \over g} \epsilon_{jkj'k'}\bigr
]\, ,\eqn\inverse$$
and  $A$ defined below \zeroone\ can be proved to have $\det A=|\tau|^6$.
The expression \almost\ is only a rewriting of \zero\ and therefore
is equivalent to \trace. 
The Poisson transformed version \almost\ is particularly useful to 
exhibit the invariance of \trace\ under modular transformations. 

The generalization of the usual $\tau \rightarrow\ -\tau^{-1}$
modular transformation that 
leaves invariant the line element
\B\   if $d\theta^1\nobreak\rightarrow d\theta^6, \,
d\theta^6 \rightarrow -d\theta^1,\,
d\theta^i \rightarrow d\theta^i,$ is
$$R_1\rightarrow R_1|\tau|,\, R_6  \rightarrow R_6|\tau|^{-1},\, \alpha
\rightarrow -{|\tau|}^{-2}\alpha, \,\beta^i\rightarrow \gamma^i,\,
\gamma^i \rightarrow -\beta^i ,\, g_{ij}\rightarrow g_{ij}\,. \eqn\modular$$

We perform the transformation \modular\ on \hham\  and identify
the result with the Poisson transformed expression \almost. 
Starting from \hham\ we write 
$$\eqalign{&{Z}(R_1|\tau|,{R_6 |\tau|^{-1}}, g_{ij},
-\alpha |\tau|^{-2},
\gamma^i, -\beta^i )\cr
&= Z'_{\rm osc}\cdot
\sum_{n_7...n_{10}} \exp\{-\pi { R_6R_1 \over 6}{\sqrt g} g^{ii'}g^{jj'}g^{kk'}
H_{ijk}H_{i'j'k'} \cr 
&\hskip90pt -{\pi\over 4|\tau|^2}{R_6\over R_1}{\sqrt g}(g^{jj'}g^{kk'} -
g^{jk'}g^{kj'}) \gamma^i \gamma^{i'}H_{ijk}H_{i'j'k'}\} \cr 
&\hskip40pt \cdot
\sum_{n_1...n_6}\exp\{-{\pi \over |\tau|^2} \bigl [{R_6\over R_1}{\sqrt g}
(g^{jj'}g^{kk'} -
g^{jk'}g^{kj'}) - i\alpha \epsilon^{jkj'k'}\bigr ]H_{1jk}H_{1j'k'}\cr
&\hskip90pt -{\pi\over 2|\tau|^2}\bigl [{R_6\over R_1}
{\sqrt g}(g^{jj'}g^{kk'} -
g^{jk'}g^{kj'}) - i\alpha \epsilon^{jkj'k'}\bigr ]
\gamma^i H_{1jk}H_{ij'k'}\cr 
&\hskip90pt
+{i\pi \over 2} \beta^iH_{1jk}\epsilon^{jkj'k'}H_{ij'k'}\}\cr}\eqn\transform$$
where $Z'_{\rm osc}$ is the transformed oscillator trace 
derived in sect.'s 4,5.
The first two lines of \transform\  and \almost\ are the
same. The rest can also be identified if one rotates the
$H_1^{jk}$ variables, defining  new variables ${\tilde H_{1jk}}$ via 
$$H_1^{jk}= {1\over 2}\epsilon^{jkj'k'}{\tilde H_{1j'k'}}.\eqn\ROT$$
Thus we have shown  that under the transformation \modular\ 
$${Z}_{\rm zero\, modes}(R_1|\tau|,{R_6 |\tau|^{-1}}, g_{ij},
-\alpha |\tau|^2, \gamma_i, -\beta_i)
= (\det A )^{{\scriptstyle{1\over 2}}}\,
{Z}_{\rm zero\, modes}(R_1, R_6, g_{ij}, \alpha, \beta_i, \gamma_i)\,.
\eqn\si$$

\chapter{The Oscillator Trace}

In this section we set up the trace calculation in detail and  find the
normal mode expansion of the `Hamiltonian' $-t{\cal H} + i2\pi\alpha
P_1 + i2\pi (\alpha\beta_i + \gamma_i)P_i$ appearing in \trace.
The result is given in (4.19) and (4.20).
It is convenient to introduce  
the dependent field strength $H_{6mn}$ given in \DEP\   and write
$$\eqalign{ -t{\cal H} + i2\pi\alpha
P_1 + i2\pi (\alpha\beta^i + \gamma^i)P_i  = &  
{i\pi\over 12}\int_0^{2\pi} d^5\theta H_{lrs}
\epsilon^{lrsmn} H_{6mn}\cr
=& {i\pi\over 2}\int_0^{2\pi} d^5\theta \sqrt{-G} 
H^{6mn} H_{6mn}\cr}\eqn\SUB$$
where 
$H^{6mn} = {1\over 2\sqrt{-G}}\epsilon^{mnlrs} H_{lrs}$ from \A .

Let's define  $\Pi^{mn}(\vec\theta,\theta^6)$,
the field conjugate to  $B_{mn}(\vec\theta,\theta^6)$, starting 
from the 6d Lagrangian
for a general (non-self-dual) two-form 
$I_6=\int d^6\theta (-{\sqrt{-G}\over 24})H_{LMN} H^{LMN}$ 
$$\Pi^{mn}\nobreak={\textstyle{\delta I_6\over \delta \partial_6
B_{mn}}}  = -{\sqrt{-G}\over 4} H^{6mn}\,.\eqn\no$$ 
In terms of $\Pi^{mn}$  \SUB\ can be written as 
$${-i\pi \int_0^{2\pi} d^5\theta 
(\Pi^{mn} H_{6mn} + H_{6mn} \Pi^{mn})}\,,\eqn\OR$$
where we have chosen a specific field ordering. 
The commutation relations of the two-form 
and its conjugate field $\Pi^{mn}(\vec\theta,\theta^6)$ are assumed
to be the standard ones 
$$\eqalign{[\Pi^{rs}(\vec\theta,\theta^6),
B_{mn}(\vec\theta',\theta^6)]
=& -i\delta^5 (\vec\theta - \vec\theta') 
(\delta^r_m\delta^s_n - \delta^r_n\delta^s_m)\cr
[\Pi^{rs}(\vec\theta,\theta^6),
\Pi^{mn}(\vec\theta',\theta^6)]
=& [B_{rs}(\vec\theta,\theta^6),
B_{mn}(\vec\theta',\theta^6)] =0\cr}\eqn\COM$$

{}From the Bianchi identity $\partial_{[L}H_{MNP]} = 0$ and the fact
that \A\ implies $\partial^L H_{LMN} =0$, it follows that
a solution to \A\ is given by a solution to the homogeneous equations

$$\partial^L\partial_L B_{MN} = 0\,;\qquad \partial^L B_{LN} =0\,.\eqn\BOX$$ 

These have a plane wave solution

$$B_{MN}(\vec\theta,\theta^6) 
=f_{MN}(p) e^{ip\cdot\theta} + (f_{MN}(p) e^{ip\cdot\theta})^\ast\eqn\PW$$

when 
$$G^{LN}p_L p_N =0\,;\qquad p^L f_{LN} =0\,.\eqn\PPW$$

Using the metric on the six-torus \B\  and solving for $p_6$ from
\PPW, we get 
$$\eqalign{ p_6 =& -{G^{6m}\over G^{66}}p_m -i {\sqrt{{G_5^{mn}\over
G^{66}}p_m p_n}}
\cr = &  -\alpha p_1 - (\alpha\beta^i + \gamma^i) p_i
-i R_6{\sqrt{G_5^{mn}p_m p_n}} \cr}\eqn\OMEGA$$
where $2\le i\le \,; 1\le m,n\le 5$. 
Now consider a wave in the 1-direction with wave vector $p_i=0$,
$p_6 = (-\alpha-i{R_6\over R_1}) p_1$, 
and use the gauge invariance of $B_{MN}$, {\it i.e.}
$f_{MN}\rightarrow f'_{MN} = f_{MN} + p_M g_N - p_N g_M$ to fix
$f'_{6n}=0$. This   gauge choice
$$B_{6n} = 0\,,\eqn\GAUGE$$ is consistent with the
commutation relations \COM\ and reduces the number of components of
$B_{MN}$ from 15 to 10. 
 
Furthermore, the $\rho\sigma j$ component of \A\ , where
$\rho,\sigma = 6,1$  and $2\le j\le 5$, can be used to 
eliminate $f_{1j}$ in terms of the six $f_{ij}$ as
$$f_{1j} = {\textstyle{(\alpha + i{R_6\over R_1}\beta^k + \gamma^k})\over
(\alpha + i{R_6\over R_1})} f_{jk} = {\textstyle p^k f_{jk}\over p^1}\,.
\eqn \FIX$$
This satisfies \PPW\ .  Finally 
the $\rho j k$ component of \A\ expresses three of the $f_{ij}$ 
in terms of the remaining three, leaving just three independent polarization
tensors corresponding to the physical degrees of freedom
of the 6d chiral two form with Spin(4) content (3,1).

We can now  expand the free quantum tensor gauge field as
$$B_{mn} (\vec\theta, \theta^6) = {\rm zero\, modes} +
\sum_{\vec p\ne 0} ( f_{mn}^\kappa b_{\vec p}^\kappa e^{ip\cdot\theta}
+ f_{mn}^{\kappa\ast} b_{\vec p}^{\kappa\dagger} e^{-ip^\ast\cdot\theta})
\eqn\NME$$ where $1\le\kappa\le 3$, $1\le m,n\le 5$, 
$p_6$ is defined by \OMEGA\ , and the sum in \NME\ is on the dual
lattice $\vec p = p_m \in {\cal Z}^5\ne \vec 0$.
Since oscillators with different polarizations commute, we can treat
each polarization separately and cube the end result. 
Also, having already computed the zero mode contribution in sect.3, we will
drop the `zero mode' term here. Thus $$B_{mn}(\vec\theta, \theta^6) = 
\sum_{\vec p\ne 0} ( b_{\vec p nm}\,e^{ip\cdot\theta}
+ b_{\vec p mn}^{\dagger} e^{-ip^\ast\cdot\theta})\,,
\eqn\SNM$$ where $b_{\vec p nm}= f_{mn}^1 b_{\vec p}^1 $, $\,$
for example. From
the mode expansion for $B_{mn}$ in \SNM, 
since $H^{6mn} = {1\over 2\sqrt{-G}}\epsilon^{mnlrs} H_{lrs}$,
we can also write one for $H^{6mn}$ as 
$$\Pi^{mn} (\vec\theta, \theta^6)  = -{\sqrt{-G}\over4} H^{6mn} =
\sum_{\vec p\ne 0} ( c^{6mn}_{\vec p}\,e^{ip^\ast\cdot\theta}
+ c^{6mn\dagger}_{\vec p}\, e^{-ip\cdot\theta})\,.
\eqn\HNM$$
Then substituting \SNM\ and \HNM\ in the first term of \OR\ we find
$$\eqalign{-i\pi \int_0^{2\pi} d^5\theta \Pi^{mn} H_{6mn}
=& -i\pi (2\pi)^5
\sum_{\vec p \ne 0} ip_6 (c_{-\vec p}^{6mn} + c_{\vec p}^{6mn\dagger})
(b_{\vec p mn} + b_{-\vec p mn}^{\dagger})\cr}\,.\eqn\MOM$$ 
We compute the commutators of these oscillator combinations
$$\eqalign{\int {d^5\theta\over(2\pi)^5}\, e^{-ip_l\theta^l} 
B_{mn}(\vec\theta, 0) 
=& \, b_{\vec p \,mn} + b_{-\vec p \,mn}^\dagger\equiv
B_{\vec p \,mn}\cr
\int {d^5\theta\over(2\pi)^5} \,e^{ip_l\theta^l}
\Pi^{mn} (\vec\theta, 0) 
=& \, c_{-\vec p}^{6mn} + c_{\vec p}^{6mn\dagger}\equiv
{\cal C}_{\vec p}^{6mn\dagger}\cr}\eqn\OSC$$
so that 
$$\eqalign{[{\cal C}_{\vec p}^{6rs\dagger}, B_{\vec {p'} \,mn}] 
=& {1\over (2\pi)^{10}} \,\int_0^{2\pi} d^5\theta d^5\theta' 
e^{i\vec p\cdot\vec\theta - i \vec{p'}\cdot\vec\theta'} 
\,[\Pi^{rs} (\vec\theta, 0), B_{mn} (\vec\theta', 0)]\cr
=& -{i\over (2\pi)^5} \delta_{\vec p, \vec p'}
(\delta_m^r\delta_n^s - \delta_n^r\delta_m^s)\,.\cr}\eqn\OSCOM$$
Normalizing ${\cal C} B \equiv {i\over{(2\pi)^5}}\tilde{\cal C} \tilde B$,
so that
$$\eqalign{[\tilde{\cal C}_{\vec p}^{6rs\dagger}, \tilde B_{\vec p' \,mn}]
=& -\delta_{\vec p, \vec p'}
(\delta_m^r\delta_n^s - \delta_n^r\delta_m^s)\cr}\eqn\NOSCOM$$
we get
$$\eqalign{-i\pi \int_0^{2\pi} d^5\theta (\Pi^{mn} H_{6mn} +
H_{6mn} \Pi^{mn} )
=& -i\pi 
\sum_{\vec p \ne 0} p_6 (\tilde{\cal C}_{\vec p}^{6mn\dagger} 
\tilde B_{\vec p mn} +
\tilde B_{\vec p mn} \tilde{\cal C}_{\vec p}^{6mn\dagger} )\,.\cr}\eqn\TOR$$
Reinserting the polarization tensors normalized as 
$f^{\kappa rs}(p) f_{rs}^\lambda (p) = \delta^{\kappa\lambda}$
and using \NOSCOM\ , we find that \TOR\ becomes
$$\eqalign{&-2 i \pi\sum_{\vec p \ne 0} p_6 
{\cal C}_{\vec p}^{\kappa\dagger} B_{\vec p}^\lambda
f^{\kappa mn}(p) f^\lambda_{mn}(p)
- i\pi \sum_{\vec p \ne 0} p_6 f^{\kappa mn}(p) f^\kappa_{mn}(p)\cr
&=  -2 i \pi\sum_{\vec p \ne 0} p_6 
{\cal C}_{\vec p}^{\kappa\dagger} B_{\vec p}^\kappa
-i \pi \sum_{\vec p \ne 0} p_6 \delta^{\kappa\kappa}\cr
&=  -2i\pi \sum_{\vec p \ne 0} p_6 
{\cal C}_{\vec p}^{\kappa\dagger} B_{\vec p}^\kappa
-\pi R_6 \sum_{\vec p} \sqrt{G_5^{mn} p_m p_n}\,\,
\delta^{\kappa\kappa}\cr}
\eqn\ORVE$$
where $1\le\kappa,\lambda\le 3$ and 
$$[{\cal C}_{\vec p}^{\kappa\dagger}, B_{\vec p'}^\lambda] =
\delta^{\kappa\lambda}\,\delta_{\vec p,\vec p'}\eqn\AOSC$$
The ordering chosen in \OR\ gives rise to the vacuum energy as the 
second term in \ORVE\ . This term is necessary for modular invariance and
requires an $SL(5,{\cal{Z}})$ invariant regularization which is
derived in Appendix A.

\chapter {The Anomaly Cancellation}

{}From \ORVE\ and \trace\ the partition function is 
$$Z = Z_{\rm zero\, modes}\,\cdot\tr e^{-2i\pi \sum_{\vec p \ne 0} p_6
{\cal C}_{\vec p}^{\kappa\dagger} B_{\vec p}^\kappa
-\pi R_6 \sum_{\vec p } \sqrt{G_5^{mn} p_m p_n}\,\,
\delta^{\kappa\kappa}}\eqn\AGAIN $$
with $p_6$ given in \OMEGA\ and $p_l = n_l\in {\cal Z}^5$ due to the torus.
We can now use the standard Fock space argument
$$tr\omega^{\sum_p p a^\dagger_p a_p} 
=\prod_p\sum_{k=o}^\infty \langle k |\omega^{p a^\dagger_p a_p} | k\rangle
=\prod_p {\textstyle 1\over {1 - \omega^p}}\eqn\OSCTR$$
 to do the trace on the oscillators in \AGAIN. The answer is
$$\eqalign{Z =& Z_{\rm zero\,modes}\cdot \bigl ( e^{-\pi R_6 \sum_{\vec n} 
\sqrt{G_5^{lm} n_l n_m}}\, \prod_{\vec n\ne \vec 0}
{\textstyle 1\over{1- e^{-i2\pi p_6}}}\bigr )^3\,.\cr}\eqn\OPF$$
\OPF\ is manifestly $SL(5,{\cal Z})$ invariant since $p_6$ is.
(In $p_6$, $\,$ $G^{6m}$ is a contravariant 5-vector defined in \Gsixinv\ .)
However, the vacuum energy $\sqrt{G_5^{lm} n_l n_m}$ is a divergent
sum. In (A.5) in Appendix A we derive its $SL(5,Z)$
invariant regularization to obtain 

$$\eqalign{Z =& Z_{\rm zero\,modes}\cdot
\bigl ( e^{ R_6 \pi^{-3} \sum_{\vec n\ne \vec 0} {\sqrt{G_5}\over
(G_{lm}n^ln^m)^3}}\,
\prod_{\vec n\ne \vec 0}
{\textstyle 1\over{1- e^{-2\pi R_6 \sqrt{G_5^{lm}n_l n_m} + i 2\pi\alpha n_1
+ i 2\pi (\alpha\beta^i +\gamma^i)n_i}}}\bigr )^3\cr}\,\eqn\SLFIVE$$

\noindent where the sum on $\vec n$ is on the original lattice
$\vec n = n^l \in {\cal Z}^5\ne\vec 0$ and the product on $\vec n$ is on
the dual lattice $\vec n = n_l\in {\cal Z}^5\ne \vec 0$.
$Z_{\rm zero\, modes}$ is given in \zeroone\ . 
To understand how the $SL(2,{\cal Z})$ invariance of $Z$ works,
 we separate the product on
$\vec n = (n,n_{\perp})\ne \vec 0$ into a product
on (all $n$, but $n_\perp\ne\nobreak (0,0,0,0)$)  
and on ($n\ne 0$, $n_{\perp} = (0,0,0,0))$, where
$n_\perp\equiv n_i$. Then 
\SLFIVE\ becomes  $$\eqalign{Z =& Z_{\rm zero\,modes}
\cdot \bigl ( e^{\textstyle{R_6\over\pi R_1}\zeta (2)} 
\prod_{n_1\ne 0}
{\textstyle 1\over{1- e^{2\pi i (\alpha n_1 + i {\textstyle {R_6\over R_1}} 
|n_1|)}}}\bigr )^3
\cr &\cdot
\bigl ( \prod_{n_i\ne (0,0,0,0)} e^{-2\pi R_6 <H>_{n_\perp}}
{\textstyle 1\over{1- e^{-2\pi R_6 \sqrt{G_5^{lm}n_l n_m} + i 2\pi\alpha n_1
+ i 2\pi (\alpha\beta^i +\gamma^i)n_i}}}\bigr )^3\cr
=& Z_{\rm zero\,modes}\cdot 
\bigl (\eta (\tau) \bar\eta(\bar\tau)\bigr )^{-3}\cr
&\cdot \bigl ( \prod_{n_i\ne (0,0,0,0)} 
e^{-2\pi R_6 <H>_{n_\perp}} 
\prod_{n_1\in {\cal Z}}
{\textstyle 1\over{1- e^{-2\pi R_6 \sqrt{G_5^{lm}n_l n_m} + i 2\pi\alpha n_1
+ i 2\pi (\alpha\beta^i +\gamma^i)n_i}}}\bigr )^3\cr}\eqn\SLTWO$$
\noindent where $\tau\equiv \alpha + i{\textstyle {R_6\over R_1}}$ and
$<H>_{n_\perp}$ is given in (A.11).

\noindent In \SLTWO\ we have separated the contribution of the `2d
massless' scalars from the contribution of the
`2d massive' scalars. The former are the modes with zero momentum
$n_{\perp}=0$ in
the transverse direction $i=2...5$, which appear  as massless bosons
on the 2-torus in the directions $1$ and $6$.  
Instead, the modes associated with $n_{\perp} \ne 0$ correspond to
massive bosons on the 2-torus. Their partition function at fixed $n_{\perp}$ 
is $$e^{-2\pi R_6 <H>_{n_\perp}} \prod_{n_1\in {\cal Z}}
{\textstyle 1\over{1- e^{-2\pi R_6 \sqrt{G_5^{lm}n_l n_m} + i 2\pi\alpha n_1
+ i 2\pi (\alpha\beta^i +\gamma^i)n_i}}}\eqn\MB$$
and is $SL(2,{\cal Z})$ symmetric by itself, since there is no anomaly
for massive states.
We show in Appendix B  how
\MB\ can be derived from the path integral 
for a complex scalar field coupled to a constant gauge field
on the 2-torus.
The modular
invariance under \modular\ reduces on the 2-torus to the standard  
$\tau \rightarrow -{1\over \tau}$ transformation plus gauge invariance. 
In this path integral derivation the  
invariance of \MB\ under \modular\  then follows by construction.

The only piece of \SLTWO\ that has an
$SL(2,{\cal Z})$ anomaly is the one associated with the `2d
massless' modes 
$$e^{\textstyle{R_6\over\pi R_1}\zeta (2)}
\prod_{n_1\ne 0}
{\textstyle 1\over{1- e^{2\pi i (\alpha n_1 + i {\textstyle {R_6\over R_1}}
|n_1|)}}} = \bigl(\eta (\tau) \bar\eta(\bar\tau)\bigr )^{-1}$$
where the Dedekind eta function 
$\eta (\tau) \equiv 
e^{\pi i\tau\over{12}}\prod_{n=1}^\infty (1 - e^{2\pi i \tau n)}$,$\,$ 
and the Riemann zeta function $\zeta(2)={\textstyle{\pi^2\over 6}}$
results from Appendix A. 
Under the $SL(2,{\cal Z})$ transformation \modular\ of sect. 3,
$\tau\rightarrow -{1\over \tau}$ and  
$$(\eta (\tau) \bar\eta(\bar\tau))^{-3}\rightarrow 
|\tau|^{-3} (\eta (\tau) \bar\eta(\bar\tau))^{-3}\,. \eqn\cinque$$

This is how the oscillator anomaly cancels the zero mode anomaly
in \si. Hence the combination $Z_{\rm zero\, modes} \cdot
(\eta (\tau) \bar\eta(\bar\tau))^{-3}$ is $SL(2,{\cal Z})$ invariant. 
 
In analogy with the modular group $SL(2,{\cal Z})$ which can be generated by
two transformations such as $\tau\rightarrow \tau + 1$ and 
$\tau\rightarrow -{\textstyle 1\over\tau}$, the mapping class groups
of the $n$-torus, {\it i.e.}
the modular groups $SL(n,{\cal Z})$ can be generated
by just two transformations as well [\Coxeter].
In Appendix C we show how the 
$SL(5,{\cal Z})$ invariance of \SLFIVE\
and the $SL(2,{\cal Z})$ invariance of \SLTWO\ imply
symmetry under the $SL(6,{\cal Z})$ generators. 

\chapter{Conclusions}
In this paper we have computed explicitly the partition function of
the M-theory fivebrane chiral two-form on a six-torus.
In analogy with the  $SL(2,{\cal Z})$ modular invariance of 2d string theory, 
the fivebrane partition function  on $T^6$  has  $SL(6, {\cal
Z})$  invariance.  We have shown this in two
steps. We started with a manifestly $SL(5, {\cal Z})$  invariant
formalism (where 5 here refers to  the directions 1...5) 
and made sure that the regularization of the vacuum energy
did not spoil it. The crucial step, however, was to prove 
an additional $SL(2, {\cal Z})$ 
symmetry in the directions 1 and 6. This $SL(2, {\cal Z})$ invariance was
achieved by the cancellation of the anomaly from the 
sum over the zero modes with the anomaly of the 
`2d massless' modes from the oscillator trace. We then showed how
the combination of these symmetries implies $SL(6, {\cal Z})$ invariance. 
 
One of the problems with the M-theory fivebrane is that it is hard to
write down a manifestly covariant Lagrangian. Nonetheless, 
our result proves  that  the M-theory fivebrane chiral two-form 
can be consistently quantized on a six-torus.

Finally, we have shown that in this case
the partition function for the fivebrane two-form has no
dependence on the spin structure. Our result depends on the
fact that we compactify on $T^2\times T^4$ and would not
hold automatically on other spaces.

\vskip20pt
\leftline{\bf Acknowledgement}

We thank Edward Witten for valuable discussions, and
L.D. thanks the IAS for its hospitality. 
\vfill\eject

\Appendix{A}
\noindent $SL(5,Z)$ INVARIANT REGULARIZATION OF THE VACUUM ENERGY

The vacuum energy $<H>\equiv {1\over 2} \sum_{\vec p\in {\cal Z}^5}
\sqrt {{G_5}^{lm}
p_lp_m} = \half \sum_{\vec p\in {\cal Z}^5}|{\vec p}|$  appearing  in sect.3 is
a divergent sum and
needs to be regularized in an $SL(5,{\cal Z})$ invariant way. 
Here the sum is on the dual lattice $\vec p = p_l\in {\cal Z}^5$. 
The answer given in (A.5) is derived as follows. 
We rewrite $<H>$ as 
$$<H> = {1\over 2} \sum_{\vec p \in {\cal Z}^5}
{\sqrt {{G_5}^{lm}p_lp_m}} = {1\over 2}
\sum_{\vec p\in {\cal Z}^5}|\vec p| 
e^{i{\vec p} \cdot {\vec x}}\,\Big|_{{\vec x} =0}\,.
\eqn\aone $$ 
We  express $|\vec p|$ in terms of its 5d Fourier transform as 
$${|\vec p|} = \int d^5y e^{-i{\vec p}\cdot
{\vec x}} (-{2\over \pi^3} \sqrt{G_5}) \,{1\over |{\vec y}|^6}\,. \eqn\star$$
Then 
$$\eqalign{&\sum_{\vec p}|\vec p| e^{i{\vec p} \cdot {\vec x}}\cr &
= -{2\over \pi^3} \sqrt{G_5} \int d^5y  {1\over |{\vec y}|^6}
\sum_{\vec p}e^{-i{\vec p}\cdot  ({\vec x} - {\vec y})} \cr
&= -{2\over \pi^3} \sqrt{G_5} \int d^5y  {1\over |{\vec y}|^6}
(2\pi)^5\sum_{{\vec n}\ne 0} \delta^5({\vec x}-{\vec y} +2\pi {\vec
n})\cr & = - 64 \pi^2  \sqrt{G_5} \sum_{{\vec n}\ne 0}
{1\over |{\vec x} + 2\pi{\vec n}|^6}\cr } \eqn\stop$$
where we have used the equality
$$\sum_{\vec p}e^{-i{\vec p}\cdot {\vec x}} = (2\pi)^5
\sum_{\vec n} \delta^5({\vec x} +2\pi {\vec n}) \eqn\stim$$
and the sum on $\vec n$ is on the original lattice 
$\vec n = n^l\in {\cal Z}^5$.
Our regularization consists in removing the ${\vec n}=0$ term from this sum.
The regularized  vacuum energy  follows from \aone\ and \stop\
$$<H> = -{1\over 2\pi^4}{\sqrt G_5} \sum_{\vec n \ne 0} {1\over
(G_{lm}n^ln^m)^3} = -32\pi^2\sqrt{G_5}\sum_{{\vec n}\ne 0} {1\over
|2\pi {\vec n}|^6}\,. \eqn\still $$
We want to show now that in the
case of zero transverse momentum, {\it i.e.} $ n_{\perp} = (n_2, n_3, n_4, n_5)
=(0,0,0,0)$,  the regularization \still\
reduces to the usual $\zeta$ function regularization.
This  is essential to give the Dedekind eta
function in sect.5 for the anomaly cancellation with the
zero modes of sect.3.

To this purpose we rewrite the vacuum energy \still\ as a sum 
(on the dual lattice $p_\perp = p_{\perp i}\in {\cal Z}^4$) of terms
at fixed transverse momentum: 
$$<H> = -32\pi^2\sqrt{G_5}\sum_{p_{\perp}}{1\over (2\pi)^4}\int
d^4z_{\perp} e^{-ip_{\perp}\cdot 
z_{\perp}} \sum_{{\vec n}\ne 0} {1\over |2\pi\vec n + z_{\perp}|^6}\,.
\eqn\stew$$
The identity \stew\ can be checked from \still\ by using \stim.
By a change of variables $ z_{\perp} \rightarrow y_{\perp} =z_{\perp}
+ 2\pi {\vec n}$,  \stew\ can be now rewritten as
$$<H> = -32\pi^2\sqrt{G_5}\sum_{p_{\perp}}{1\over (2\pi)^4}\int d^4y_{\perp}
e^{-ip_{\perp}\cdot
y_{\perp}} \sum_{n^1\in{\cal Z}\ne 0} {1\over |2\pi n^1 + y_{\perp}|^6}
\eqn\newstew $$
where $|2\pi n^1 + y_{\perp}|^2\equiv 
[(2\pi n^1)^2 G_{11} + 2 (2\pi n^1) G_{1i} y_\perp^i + 
y_\perp^i y_\perp^j G_{ij}]$.
Now we consider the $p_{\perp} =0$ part only in \newstew\ and do  
the 4d integration in $d^4y_{\perp}$ with the result 
$$<H> = -32\pi^2\sqrt{G_5} {1\over (2\pi)^4} \sum_{n\in {\cal Z}}
{\pi^2\over g}{1\over
(2\pi n)^2}{1\over 2{R_1}^2}\,.\eqn\stall$$ 
Remembering that ${\sqrt{G_5}\over\sqrt{g}}= R_1$ and that
$\sum^\infty_{n=1} {1\over n^2}=
\zeta (2)={\pi^2 \over 6}$, we finally obtain 
$$<H>_{p_{\perp}=0} = -{1\over 12}{1\over R_1} = {1\over R_1}
\zeta (-1) .\eqn\stay$$
\vfill\eject

We note that computing the $d^4y_{\perp}$ integration in \newstew\ for
all $p_\perp$, we recover the spherical Bessel functions which appear
for massive bosons:
$$<H> = \sum_{p_\perp} <H>_{p_\perp}\eqn\PERP$$
where 
$$<H>_{p_\perp} = - |p_\perp |^2 R_1 \sum_{n^1=1}^\infty 
{\rm cos}(p_\perp\cdot\beta 2\pi n^1)
[ K_2(2\pi n^1 R_1 |p_\perp |) - K_0(2\pi n^1 R_1 |p_\perp |) ] \eqn\BES $$

For $\beta^i = 0$, \PERP\ can be expressed in terms of a standard
integral occurring in the effective potential for a 4d scalar field at
finite temperature:
$$<H> = - {1\over 2\pi^2 R_1} \sum_{p_\perp} \int_0^\infty dx 
{\textstyle x^2\over {\sqrt{x^2+a^2}}}
{\textstyle 1\over {(e^{\sqrt{x^2 + a^2}} -1})}\eqn\FITMP$$
where $a= 2\pi R_1 |p_\perp|$.

\vfill\eject

\Appendix{B}
\noindent{$SL(2,{\cal Z})$ SYMMETRIC PARTITION FUNCTION OF A 2D
MASSIVE SCALAR FIELD}

In (5.5)
we wrote the contribution from the oscillator trace to 
the  partition function of the chiral two-form on the torus as a
product over values of the transverse
momentum $n_\perp$. We show here that  
each term in the product with fixed $n_{\perp}\ne 0$ given in 
\MB\ 
is the square root of 
the partition function on $T^2$ of a massive complex scalar 
with $m^2 \equiv g^{ij} n_i n_j$ coupled to
a constant gauge field
$A^\mu \equiv i G^{\mu i}  n_i$ with $\mu,\nu ={1,6};\,i,j = 2,\ldots 5$. 
We show that \MB\
is a gauge transformation on the 2d gauge field 
combined with an $SL(2,{\cal Z})$ transformation on $T^2$.
The metric on $T^2$ is $h_{11} = R_1^2\, , h_{66} = R_6^2 +
\alpha^2 R_1^2\, , \, h_{16} = -\alpha R_1^2$. Its inverse is
$h^{11}= G^{11}$, $h^{66}=G^{66}$ and $h^{16}=G^{16}$. 
The invariance under \modular\
follows since the 2d partition function \MB\ can 
be derived from a manifestly $SL(2,{\cal Z})$ invariant {\it path integral}
on the 2-torus: 
We start with 
$$\eqalign{ {\rm P.I.}
&=\int d\phi \,d\bar\phi \,\,
e^{-\int_0^{2\pi} d\theta^1 \int_0^{2\pi} d\theta^6\,\,
h^{\mu\nu}(\partial_\mu + A_\mu )\bar\phi
(\partial_\nu - A_\nu) \phi + m^2 \bar\phi \phi}\cr
&=\int d\phi \,d\bar\phi \, 
e^{-\int_0^{2\pi} d\theta^1 \int_0^{2\pi} d\theta^6
\bar\phi ( -h^{\mu\nu}\partial_\mu
\partial_\nu + 2A^\mu\partial_\mu + G^{ij} n_i n_j ) \phi}\cr 
&=\int d\bar\phi\, d\phi 
e^{-\int_0^{2\pi} d\theta^1 \int_0^{2\pi} d\theta^6
\bar\phi (-(({1\over R^1})^2 + ({\alpha\over R_6})^2)\partial_1^2 
-({1\over R_6})^2\partial_6^2 -2 {\alpha\over {R_6^2}}
\partial_1\partial_6 + 2A^1\partial_1 + 2A^6\partial_6
+ G^{ij} n_i n_j )\phi}\cr  
& = \det \Bigl ( 
[- ( {\scriptstyle{1\over R_1^2}} + ({\scriptstyle {\alpha\over R_6}})^2 )
\partial_1^2 - ({\scriptstyle {1\over R_6}})^2\partial_6^2 
- 2\alpha ({\scriptstyle {1\over R_6}})^2\partial_1\partial_6  
+ G^{ij} n_i n_j  + 2 G^{1i}n_i \partial_1 
+ 2G^{6i}n_i \partial_6 ]\,\Bigr )^{-1}\cr
&= e^{- \tr \ln \Bigl [ 
- ( {\scriptstyle{1\over R_1^2}} + ({\scriptstyle {\alpha\over R_6}})^2 )
\partial_1^2 - ({\scriptstyle {1\over R_6}})^2\partial_6^2
- 2\alpha ({\scriptstyle {1\over R_6}})^2\partial_1\partial_6
+ G^{ij} n_i n_j  + 2 G^{1i}n_i \partial_1 
+ 2G^{6i}n_i \partial_6 \, \Bigr ] }\cr} \eqn\Buno$$

The trace on the momentum eigenfunctions of \Buno\ reduces to the sum
$$ {\rm P. I.} =
e^{-\sum_{s\in{\cal Z}}\sum_{r\in {\cal Z}} \Bigl [ \ln 
({4\pi^2 \over \beta^2}r^2 +  
( {\scriptstyle{1\over R_1^2}} + ({\scriptstyle {\alpha\over R_6}})^2 ) s^2
+ 2\alpha ({\scriptstyle {1\over R_6}})^2 r s 
+ G^{ij} n_i n_j  + 2 G^{1i}n_i s + 2G^{6i}n_i r )
\, \Bigr ]}
\eqn\btwo$$
where $\beta \equiv 2\pi R_6$. 
To evaluate the divergent sum on $r$,
we define $E^2 \equiv G^{lm}_5 n_l n_m$ where $n_1\equiv s$ and 
$$\eqalign{\nu (E) &= \sum_{r\in {\cal Z}} \ln({4\pi^2 \over \beta^2}r^2 
+ ( {\scriptstyle{1\over R_1^2}} + ({\scriptstyle {\alpha\over R_6}})^2 ) s^2
+ 2\alpha ({\scriptstyle {1\over R_6}})^2 r s
+ G^{ij} n_i n_j + 2 G^{1i}n_i s + 2G^{6i}n_i r)\cr
&= \sum_{r\in {\cal Z}} \ln \Bigl [{4\pi^2 \over \beta^2} 
(r + \alpha s + (\alpha\beta^i + \gamma^i) n_i)^2 
+ E^2 \Bigr ]\,.\cr}\,. 
\eqn\bthree$$
Then 
$$\eqalign{{\partial \nu (E)\over \partial E} & = 
\sum_r {2E\over {4\pi^2\over \beta^2 } (r + \alpha s + 
(\alpha\beta^i +\gamma^i) n_i )^2 
+ E^2} \cr
& = {\beta \sinh{\beta E}\over {\cosh{\beta E} - 
\cos{2\pi (\alpha s + (\alpha\beta^i +\gamma^i) n_i) }}}\cr
& = \partial_E\ln \Bigl [ \cosh{\beta E} - 
\cos{2\pi \bigl ( \alpha s + (\alpha\beta^i +\gamma^i) n_i \bigr)} \Bigr ]\cr}
\eqn\bfour$$
where we have used the fact that $\sum_{n\in {\cal Z}} 
{2y\over {(2\pi n + z)^2 +y^2}} = {\sinh{y}\over {\cosh y -\cos z}}$.
Integrating \bfour\ we get
$$\nu (E) = \ln \bigl [ \cosh{\beta E} - 
\cos{2\pi \bigl (\alpha s  + (\alpha\beta^i +\gamma^i) n_i\bigr )} \bigr ]
+ \ln 2 \eqn\bfive$$
where the integration constant in \bfive\
ensures an $SL(2,{\cal Z})$ invariant regularization of \btwo. 
Consequently for $n_1\equiv s$, we have that \Buno\ is 
$$\eqalign{({\rm P.I.})^{\half} &=  
\prod_{s\in {\cal Z}} {1\over 
\sqrt 2\sqrt {\cosh{\beta E} - 
\cos{2\pi ( \alpha s + (\alpha\beta^i +\gamma^i) n_i )}}}\cr
&= \prod_{s\in {\cal Z}} {e^{-{\beta E\over 2}}\over
{1 - e^{-\beta E + 2\pi i ( \alpha s + (\alpha\beta^i +\gamma^i) n_i )}}}\cr
&= e^{-\pi R_6 {\sum_{s\in {\cal Z}} \sqrt{G_5^{lm} n_l n_m}}}
\prod_{s\in {\cal Z}}
{1\over 
{1 - e^{-2\pi R_6 \sqrt{G_5^{lm} n_l n_m} 
+ 2\pi i  \alpha s 
+ 2\pi i (\alpha\beta^i + \gamma^i) n_i}}}\,.\cr
&= e^{-2\pi R_6 <H>_{n\perp}}
\prod_{n_1\in {\cal Z}}
{1\over
{1 - e^{-2\pi R_6 \sqrt{G_5^{lm} n_l n_m} 
+ 2\pi i\alpha n_1 
+ 2\pi i (\alpha\beta^i + \gamma^i) n_i}}}\,.\cr}
\eqn\bfinal$$
where $<H>_{n\perp}$ is a sum over spherical Bessel functions given in \BES\ . 
It is not at all obvious
that ${\rm P.I.}$ given by this formula is invariant under 
$\tau\rightarrow -\tau^{-1}$, where $\tau \equiv \alpha + i{R_6\over R_1}$
and $G_5^{lm}$ is given in \Gfiveinv\ , but it is true by construction.  
\vfill\eject
Furthermore \bfinal\ is invariant under the transformation \modular\ , which
also changes the 6d metric parameters $\beta^i$ and $\gamma^i$. 
Since $A_\mu\equiv h_{\mu\nu} i n_i G^{\nu i}$ where $\mu,\nu 
= {1,6}$, the transformation \modular\ on $A^\mu$ corresponds to
a gauge transformation $A^\mu\rightarrow A^\mu + \partial^\mu\lambda $,
and $ \phi \rightarrow e^{i\lambda}$, $\bar\phi\rightarrow e^{-i\lambda}$
where $$\eqalign{\lambda (\theta^1, \theta^6)
&= \theta^1 \Bigl [ {i\gamma^i\over {R_1^2 |\tau|^2}}
+ i {\alpha\over R_6^2} ({\alpha\gamma^i\over |\tau|^2} + \beta^i) \Bigr ]
- \theta^6 \, i {\alpha\gamma^i\over |\tau|^2}\cr}\,. \eqn\LAM $$
Hence \bfinal\ and thus \MB\ are invariant under \modular\ .

\vfill\eject

\Appendix{C}
\noindent GENERATORS OF $SL(n,{\cal Z})$

The $SL(n,{\cal Z})$ unimodular groups can each be generated by two
matrices[\Coxeter]. For $SL(6,{\cal Z})$ these can be chosen to be
$$U_1 = \left (\matrix{0&1&0&0&0&0\cr
0&0&1&0&0&0\cr
0&0&0&1&0&0\cr
0&0&0&0&1&0\cr
0&0&0&0&0&1\cr
1&0&0&0&0&0\cr}\right)\,;\qquad
U_2 = \left (\matrix{1&0&0&0&0&0\cr
1&1&0&0&0&0\cr
0&0&1&0&0&0\cr
0&0&0&1&0&0\cr
0&1&0&0&1&0\cr
0&0&0&0&0&1\cr}\right)\,.\eqn\U$$
That is every matrix $M$ in
$SL(n,{\cal Z})$ can be written 
as a product $U_1^{n_1} U_2^{n_2} U_1^{n_3}\dots$.
The matrices $U_1$ and $U_2$ act on the basis vectors of the
six-torus $\vec\alpha_I$ where $\vec\alpha_I\cdot\vec\alpha_J
= G_{IJ}$. For our metric \Gsix\ , the transformation
$$\left(\matrix{\vec\alpha'_1\cr
\vec\alpha'_2\cr
\vec\alpha'_3\cr
\vec\alpha'_4\cr
\vec\alpha'_5\cr
\vec\alpha'_6\cr}\right)
=U_2\left(\matrix{\vec\alpha_1\cr
\vec\alpha_2\cr
\vec\alpha_3\cr
\vec\alpha_4\cr
\vec\alpha_5\cr
\vec\alpha_6\cr}\right)\eqn\TPLUSONE$$
corresponds to
$$R_1\rightarrow R_1 , R_6  \rightarrow R_6, \alpha
\rightarrow \alpha - 1, \beta^i\rightarrow \beta^i,
\gamma^i \rightarrow \gamma^i+\beta^i, g_{ij}\rightarrow
g_{ij}\eqn\TPOMET$$
which leaves invariant the line element \B\
if $d\theta^1\nobreak\rightarrow d\theta^1 - d\theta^6, \,
d\theta^6 \rightarrow d\theta^6,\,
d\theta^i \rightarrow d\theta^i.$
$U_2$ is the generalization of the usual $\tau\rightarrow\tau - 1$
modular transformation.
It is easily checked that $U_2$ is an invariance of the partition function
\SLFIVE\ and \zeroone.
The less trivial generator $U_1$ can be related to the tranformation
\modular\ that we study in the text as follows:
$$U_1 = U' M_5\eqn\TONEOVERT$$
where $M_5$ is an $SL(5,{\cal Z})$ transformation given by
$$M_5 = \left (\matrix{0&0&-1&0&0&0\cr
0&1&0&0&0&0\cr
0&0&0&1&0&0\cr
0&0&0&0&1&0\cr
0&0&0&0&0&1\cr
1&0&0&0&0&0\cr}\right)\eqn\Mfive$$
and
$U'$ is the matrix corresponding to \modular\ :
$$U' = \left (\matrix{0&1&0&0&0&0\cr
-1&0&0&0&0&0\cr
0&0&1&0&0&0\cr
0&0&0&1&0&0\cr
0&0&0&0&1&0\cr
0&0&0&0&0&1\cr}\right)\,.\eqn\UMOD$$
Hence the $SL(5,{\cal Z})$ symmetry of \SLTWO\ together with its
invariance under the modular transformation \modular\ implies via \TONEOVERT\
invariance under the $SL(6,{\cal Z})$ generator $U_1$.
So due to its symmetry under both generators, the
partition function is invariant under the modular group
$SL(6,{\cal Z})$, the mapping class group of the six-torus.

\refout
\end